\documentclass[sigconf, nonacm]{acmart}
\newcommand\vldbdoi{XX.XX/XXX.XX}

\newcommand\vldbpagestyle{plain} 

\usepackage{enumerate}
\usepackage[toc,page]{appendix}
\usepackage{color,tabularx,graphicx,amsmath,float,subfig}
\usepackage{listings}
\usepackage{pythonhighlight}
\usepackage{multirow}
\usepackage{multicol}
\usepackage{cleveref}
\usepackage{enumitem}
\usepackage{lipsum}
\usepackage{multirow}
\usepackage{caption}
\usepackage{xcolor,colortbl}
\usepackage{array}
\usepackage{xspace}
\usepackage{amsthm}

\theoremstyle{definition}
\newtheorem*{definition}{Definition}

\usepackage[justification=centering]{caption}
\newcolumntype{H}{>{\setbox0=\hbox\bgroup}c<{\egroup}@{}}

\crefformat{section}{\S#2#1#3} 
\crefformat{subsection}{\S#2#1#3}
\crefformat{subsubsection}{\S#2#1#3}

\definecolor{Gray}{gray}{0.85}
\usepackage{bbm}

\AtBeginDocument{%
  \providecommand\BibTeX{{%
    \normalfont B\kern-0.5em{\scshape i\kern-0.25em b}\kern-0.8em\TeX}}}

\newcommand*{\eg}{e.g.,\xspace}
\newcommand*{\ie}{i.e.,\xspace}

\newcommand*{\etal}{\emph{et~al.}\xspace}
\newcommand*{\synrd}{\textsf{SynRD}\xspace}

\begin{document}
\title{Epistemic Parity: Reproducibility as an Evaluation Metric for Differential Privacy [Experiment, Analysis \& Benchmark]}

\author{Lucas Rosenblatt}\authornote{Rosenblatt is the first author, Howe and Stoyanovich are the senior authors, other authors are listed alphabetically.}
\affiliation{%
 \institution{New York University}
  \city{New York, NY}
  \state{USA}
}
\email{lucas.rosenblatt@nyu.edu}

\author{Bernease Herman}
\affiliation{%
  \institution{University of Washington}
  \city{Seattle, WA}
  \country{USA}
}
\email{bernease@uw.edu}

\author{Anastasia Holovenko}
\affiliation{%
 \institution{Ukrainian Catholic University}
  \city{Lviv}
  \country{Ukraine}
}
\email{anastasia.holovenko@ucu.edu.ua}

\author{Wonkwon Lee}
\affiliation{%
\institution{New York University}
  \city{New York, NY}
  \country{USA}
}
\email{wl2733@nyu.edu}

\author{Joshua Loftus}
\affiliation{%
  \institution{London School of Economics}
  \city{London}
  \country{UK}
}
\email{J.R.Loftus@lse.ac.uk}

\author{Elizabeth McKinnie}
\affiliation{%
  \institution{Microsoft}
  \city{Seattle, WA}
  \country{USA}
}
\email{Elizabeth.McKinnie@microsoft.com}

\author{Taras Rumezhak}
\affiliation{%
 \institution{Ukrainian Catholic University}
  \city{Lviv}
  \country{Ukraine}
}
\email{rumezhak@ucu.edu.ua}

\author{Andrii Stadnik}
\affiliation{%
 \institution{Ukrainian Catholic University}
  \city{Lviv}
  \country{Ukraine}
}
\email{andrii.stadnik@ucu.edu.ua}

\author{Bill Howe}
\affiliation{%
 \institution{University of Washington}
  \city{Seattle, WA}
  \country{USA}
}
\email{billhowe@uw.edu}

\author{Julia Stoyanovich}
\affiliation{%
  \institution{New York University}
  \city{New York, NY}
  \state{USA}
}
\email{stoyanovich@nyu.edu}

\begin{abstract}
Differential privacy (DP) data synthesizers are increasingly proposed to afford public release of sensitive information, offering theoretical guarantees for privacy (and, in some cases, utility), but limited empirical evidence of utility in practical settings.  Utility is typically measured as the error on representative proxy tasks, such as descriptive statistics, multivariate correlations, the accuracy of trained classifiers, or performance over a query workload. The ability for these results to generalize to practitioners' experience has been questioned in a number of settings, including the U.S. Census.  In this paper, we propose an evaluation methodology for synthetic data that avoids assumptions about the representativeness of proxy tasks, instead measuring the likelihood that published conclusions would change had the authors used synthetic data, a condition we call epistemic parity. Our methodology consists of reproducing empirical conclusions of peer-reviewed  papers on real, publicly available data, then re-running these experiments a second time on DP synthetic data and comparing the results.

We instantiate our methodology over a benchmark of recent peer-reviewed papers that analyze public datasets in the ICPSR social science repository. We model quantitative claims computationally to automate the experimental workflow, and model qualitative claims by reproducing visualizations and comparing the results manually.  We then generate DP synthetic datasets using multiple state-of-the-art mechanisms, and estimate the likelihood that these conclusions will hold.  We find that, for reasonable privacy regimes, state-of-the-art DP synthesizers are able to achieve high epistemic parity for several papers in our benchmark.  However, some papers, and particularly some specific findings, are difficult to reproduce for any of the synthesizers.  Given these results, we advocate for a new class of mechanisms that can reorder the priorities for DP data synthesis: favor stronger guarantees for utility (as measured by epistemic parity) and offer privacy protection with a focus on application-specific threat models and risk-assessment.
\end{abstract}

\maketitle

\pagestyle{\vldbpagestyle}

\section{Introduction}
\label{sec:intro}

Differential privacy (DP) has been studied intensely for over a decade, and has recently enjoyed uptake in both the private and public sectors.  In situations where the downstream analysis is known, one can design specialized mechanisms with high utility~\cite{mckenna2021winning,mckenna2022aim}.  But an active research area is to design general DP data synthesizers (henceforth, synthesizers) that model the entire data distribution, inject noise, then sample the noisy model to generate synthetic datasets intended to be broadly usable in a variety of unanticipated applications. Evidence to support claims of general utility is typically presented as results on proxy tasks over common public datasets (e.g., the ubiquitous Adult dataset~\cite{adult}).  Proxy tasks may include descriptive statistics, queries involving one or two variables~\cite{hill2015evaluating,hay2016principled,takagi2021p3gm,tao2021benchmarking}, classification accuracy~\cite{ding2020differentially,takagi2021p3gm,zhang2017privbayes}, and information theoretic measures~\cite{zhang2017privbayes}. Although these proxy tasks are procedurally representative of real tasks, the implicit claim of generalization to practice is rarely explored.  

Limited empirical evidence on relevant tasks undermines trust in the practical use of DP.  The US Census Bureau adopted DP for disclosure avoidance in the 2020 census, interpreting federal law (the Census Act, 13 U.S.C. § 214, and the Confidential Information Protection and Statistical Efficiency Act of 2002) as a mandate to use advanced methods to protect against computational reconstruction attacks unforeseen when the laws were passed.  
But the adoption of DP for the Census was met with resistance among many in the research community, who contend that data infused with DP noise affects demographic totals~\cite{ruggles2019differential} and exacerbates underrepresentation of minorities~\cite{kenny2021use, ganev2021robin}.  Besides the research implications, there are potential consequences for policy: Block grants are allocated based on minority populations as measured by the census data, and underrepresentation can lead to underfunding integral services including Medicaid, Head Start, SNAP, Section 8 Housing vouchers, Pell Grants, and more~\cite{christ2022differential}.  Although the Census Bureau held workshops, released demonstration datasets, and published technical reports to support the community, these outreach efforts realized limited success; multiple lawsuits are still pending as of May 2023.

Despite these challenges, DP still offers stronger guarantees of disclosure protection than, and similar utility to, alternative proposals (\eg k-anonymity, swapping~\cite{christ2022differential}).  DP, when used correctly, ensures that any inferences conducted on data do not reveal whether a single individual's information (including, for example, their gender or race) was included in the data for analysis \cite{dwork2006calibrating}.  DP can therefore not only protect privacy, but also enable access to protected demographic attributes necessary for research on fairness and equity in machine learning~\cite{jagielski2019differentially}. 

\paragraph{Characterizing DP Error} 
A practical method of operationalizing DP is to learn a (noise-infused) model of a dataset, then sample that privatized model to generate synthetic data that can be released publicly~\cite{dwork2009complexity,hardt2010simple,DBLP:journals/corr/abs-2001-09700,rosenblatt2020differentially,vietri2020new,mckenna2021winning}.  Ideally, this approach would provide a drop-in replacement for the original data that can be used in \emph{any} downstream context to produce reasonably faithful results with strong privacy guarantees.  But this ideal is unrealizable, both theoretically and practically.  Overly accurate estimates of too many statistics are blatantly non-private, affording full reconstruction of the original dataset~\cite{DBLP:conf/pods/DinurN03}.  For any DP synthetic dataset, some statistics will tend to be faithful to the original data, while others will incur essentially arbitrary error.  If the privacy budget is allocated uniformly across features, descriptive statistics of each individual feature will be faithful, but the conditional probabilities and marginals needed to construct the joint probability distribution, which is needed for general inference, will be unreliably noisy, and vice versa.
Utility loss may also be non-uniform across subsets of a dataset, in some cases exacerbating inequity and leading to underrepresentation~\cite{bagdasaryan2019differential, kenny2021use} or to error rate disparities~\cite{DBLP:journals/corr/abs-2204-12903}.  Designers of DP synthesizers must therefore make some kind of educated guess about which tasks should be preserved and which can be ignored. Further, the error introduced by DP methods can and should be incorporated into statistical models explicitly, just as other sources of error are modeled explicitly. However, current DP synthesizers tend not to provide formal descriptions of the error they introduce; this lack of error guarantees is a major drawback of private data release.  Our work does not address this limitation, but does help provide an empirical motivation for doing so.

\begin{figure}[!tb]
\centering
\includegraphics[width=7cm]{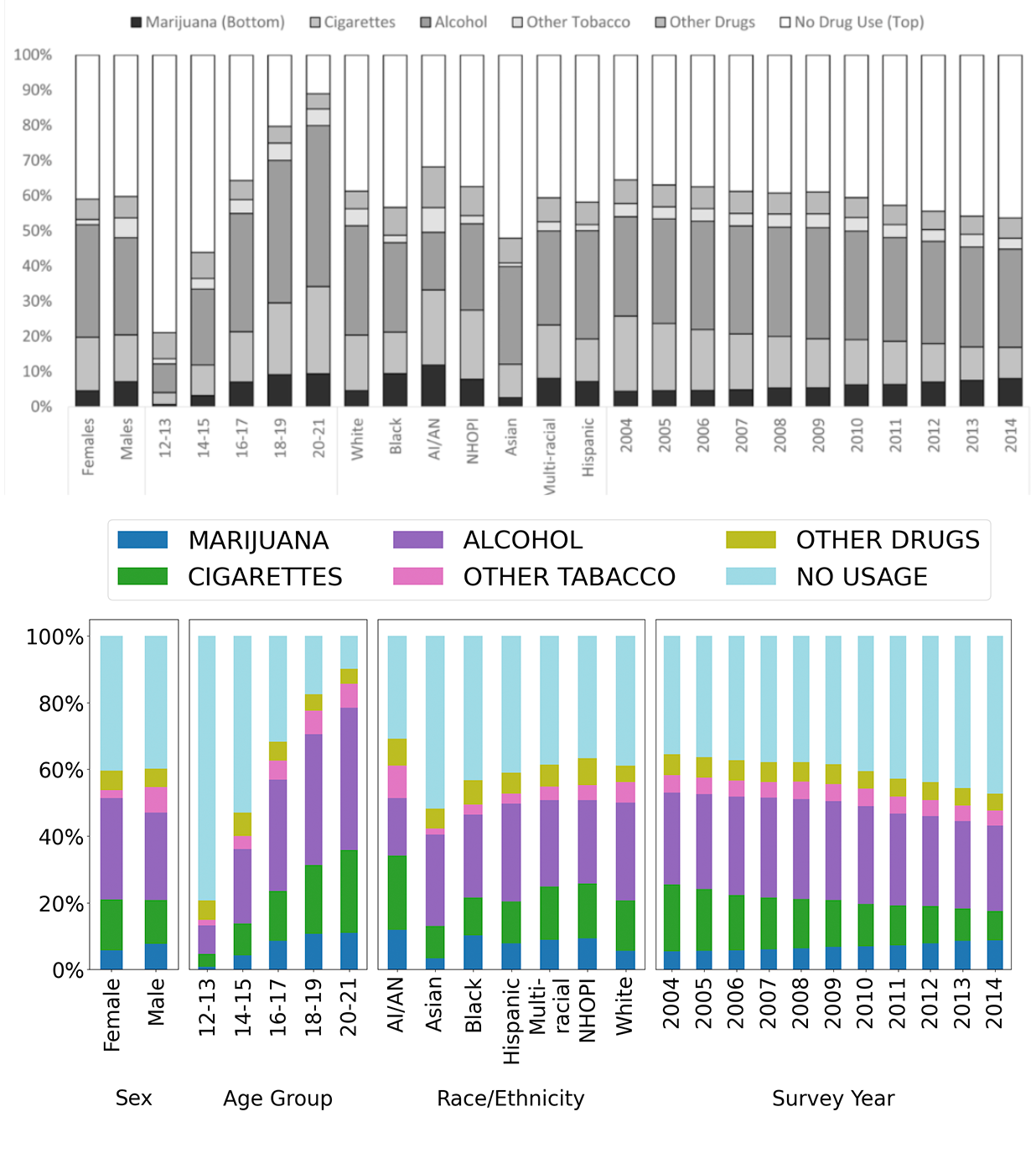}
\caption{A visual finding from~\citet{fairman2019marijuana}, describing the rate of drug use across demographic groups (top) and our qualitative reproduction under DP, using MST at $\epsilon=e$. Agreement is subjectively high, though imperfect.}
\label{fig:qualitative}
\end{figure}

\paragraph{Methodology.} 
We propose an evaluation methodology for DP synthesizers based on reproducibility: that \textit{published findings on the original dataset should be replicable on a noise-infused dataset.}  We identify conclusions in the text of published papers, extract relevant findings supporting those conclusions, implement the corresponding statistical tests using the authors' data, generate synthetic datasets using state-of-the art DP synthesizers, re-apply the statistical tests over the synthetic data, and then determine if the findings still hold.  If all findings hold, we say that the DP synthesizer achieves  \emph{epistemic parity} for that paper.  

We instantiate our methodology over a benchmark of peer-reviewed sociology papers that are based on public  data from the Inter-university Consortium for Political and Social Research (ICPSR) repository.  We model quantitative results as an inequality between two numbers, for example, ``Those using marijuana first (vs. alcohol or cigarettes first) were more likely to be Black, American Indian/Alaskan Native, multiracial, or Hispanic than White or Asian.''~\cite{fairman2019marijuana} In addition to quantitative results, we consider qualitative results: a subjective assessment of whether key visualizations in the paper expose the same relationships when recreated over synthesized data.   Figure~\ref{fig:qualitative} gives an example, presenting a visualization from~\citet{fairman2019marijuana}.  In this figure, we show the break-down of drug use across demographic groups, taken directly from the publication (top of the figure), and compare it with a reproduction of the plot based on DP synthetic data generated with the MST synthesizer at $\epsilon = e$ (bottom of the figure).  

Following~\citet{errington2021reproducibility}, and as is common in the reproducibility literature, our aim was not to reproduce every finding in every paper; rather it was to identify and reproduce a selection of key findings from each paper.  For generality, interpretability and simplicity, we consider whether a conclusion holds over synthetic data to be true if the the two quantities are in the same relative order, and do not attempt to measure the change in effect size or the statistical significance of the difference between the original and synthetic result.  

\paragraph{Benchmark and results.} ICPSR is an NSF-funded repository for social science data holding over 100,000 publications associated with 17,312 studies.  A study typically involves hundreds of variables and supports dozens of papers. Each paper can be considered to be deriving its own dataset (selected variables and selected rows) from the source data of the study. We apply DP methods to synthesize data for these paper-specific, study-derived datasets.  

 ICPSR studies are publicly available by policy, which enables us to instantiate the epistemic parity methodology and develop a benchmark.  
Notably, there is increasing demand from the ICPSR leadership and community to support keeping sensitive data private, while generating DP synthetic subsets to support reproducibility.  Our methodology can be used to respond to this demand. 

\emph{Paper selection.} The benchmark we define consists of 4 datasets and eight recent peer-reviewed papers selected for impact, accessibility of the topic to non-experts, recency, and several other criteria.  We extracted findings and attempted to reproduce them, following the "same data, different code, different team" approach to reproducibility, encountering challenges commonly reported in that literature including undocumented data versioning, unspecific or incomplete methodologies, and irreconcilable differences between our reproduction and what the authors report.
Each paper received, at a minimum, attention from two researchers with advanced degrees in either computer science, statistics, or both, and at least thirty hours of work. A complete list of papers that we attempted to reproduce, and the issues we encountered during reproduction, in our public GitHub repository, under \texttt{archive}, available at \href{https://github.com/DataResponsibly/SynRD}{https://github.com/DataResponsibly/SynRD}.

\emph{DP synthesizer selection.} We use six state-of-the-art DP synthesizers, namely, MST~\cite{mckenna2021winning}, PrivBayes~\cite{zhang2017privbayes}, PATECTGAN~\cite{rosenblatt2020differentially}, AIM~\cite{mckenna2022aim}, PrivMRF~\cite{cai2021data}, and GEM~\cite{liu2021iterative} executing each at their recommended settings. We describe these methods in Section~\ref{sec:related}.

\emph{Summary of results.} We find that marginals-based and Bayes-net based state-of-the-art DP synthesizers are able to achieve high epistemic parity for five out of eight papers in our benchmark, but that some papers, and particularly some specific findings, are difficult to reproduce for any of the synthesizers, suggesting a basis for a new benchmark.  The papers on which high epistemic parity is achieved use relatively low-dimensional tabular data. However, as we show empirically, large domain and high-dimensional settings are still a bottleneck for increased adoption of DP synthesizers.

\paragraph{Roadmap and Contributions.}  We discuss related work on reproducibility, DP synthesis and evaluation of DP  in Section~\ref{sec:related}, give background on DP in Section~\ref{sec:prelims}, and then present our contributions:

\begin{itemize}
\item We propose the epistemic parity evaluation methodology, based on reproducing qualitative and quantitative empirical findings in peer-reviewed papers over DP synthetic datasets (Section~\ref{sec:model}). 

\item We instantiate the epistemic parity methodology for eight peer-reviewed social science publications, creating a reusable benchmark for evaluating synthesizers (Section~\ref{sec:benchmark}).

\item We present \synrd, an open-source benchmarking package that automates epistemic parity evaluation, and can be extended to include additional DP synthesizers, publications, and finding types (Section~\ref{sec:software}).

\item We present experimental results on our benchmark, using five state-of-the-art DP synthesizers (Section~\ref{sec:results}). 
\end{itemize}

We conclude with a discussion of the results, identifying trade-offs and motivating a new class of privacy techniques that favor strong epistemic parity and de-emphasize  privacy risk, in Section~\ref{sec:conc}. 

\section{Related Work}
\label{sec:related}

\paragraph{DP synthesis.}  In our evaluation, we considered five state-of-the-art private data release methods: MST, AIM, PrivMRF, PATECTGAN, PrivBayes ang GEM. We acknowledge that many other methods exist for generating DP data~\cite{dwork2009complexity,hardt2010simple,DBLP:journals/corr/abs-2001-09700,vietri2020new}. We chose this set informed by recent work by~\citet{tao2021benchmarking, mckenna2022aim} showing that, over randomized query workloads on tabular data, MST, AIM and PrivMRF are the highest-performing marginal-based methods, that PrivBayes is the highest-performing Bayes-net-based method, and that PATECTGAN and GEM are the highest-performing deep learning based methods. AIM, PrivMRF and GEM are more recent than MST; they were not included in the recent dedicated DP synthesizer benchmarking survey by~\citet{tao2021benchmarking} and are currently considered to be the state-of-the-art DP synthesizers.

PrivBayes derives a Bayesian model and adds noise to all $k$-way correlations to ensure differential privacy~\cite{zhang2017privbayes}. This method was published in 2017, yet it remains competitive with and even outperforms other methods, described next.  PrivBayes has demonstrated efficacy in a number of settings, but is computationally inefficient for high-dimensional data.

MST relies on the Private-PGM graphical model to construct a maximum spanning tree among attributes in the data feature space, where edges are weighted by mutual information~\cite{mckenna2018optimizing}. By measuring 1-, 2- and some 3-way marginals, MST is able to create a high-fidelity low-dimensional approximation of the joint distribution between all attributes, leading to impressive statistical utility on metrics like mean, standard deviation, and bivariate correlations.   

AIM, like MST, relies on the Private-PGM for parameterizing the underlying distribution~\cite{mckenna2022aim}.  However, unlike MST, AIM  is \textit{workload aware}: It parameterizes a private synthetic distribution through an iterative process akin to that of the fundamental DP synthesis Multiplicative Weights Exponential Mechanism (MWEM) algorithm~\cite{hardt2010simple}, taking advantage of a set of queries of interest, specified \textit{a priori}.  In their evaluation, the authors of AIM found that it outperformed MST convincingly, especially for higher vales of $\epsilon$. 

PrivMRF is another marginal-based algorithm that relies on Private-PGM~\cite{cai2021data}.  Its novelty lies in a clever selection of marginals to measure, which aligns with three main criteria: marginals should be low-dimensional, the graph of marginals should be small, and any junction tree transform on the graph of marginals should not result in domain blowup. PrivMRF is competitive with PrivBayes and MST, as demonstrated by both~\cite{cai2021data} and~\cite{mckenna2022aim}.

PATECTGAN relies on a conditional generative adversarial network tuned to tabular data, where the discriminator has privacy constraints~\cite{xu2019modeling, rosenblatt2020differentially}. The high up-front cost of initializing and updating the weights in a full discriminator network means that PATECTGAN is limited in low dimensional settings, but shows particular promise when the data is higher-dimensional.

GEM proposes an ``Adaptive Measurements'' framework for private synthetic data algorithms. Iteratively, an algorithm (1) privately selects a set of queries; (2) obtains noisy measurements of these queries; and (3) updates an approximating distribution according to some loss function. The GEM algorithm uses a similar loss function to the RAP private data release algorithm \cite{aydore2021differentially}, although with small modifications that lead to significantly improved practical performance \cite{liu2021iterative,mckenna2022aim}.

\paragraph{Workload-aware synthesizers.}  The AIM and GEM synthesizers in our study would be considered workload-aware in that they are able to adaptively select marginals based on some \textit{a priori} specified workload of interest. In our setting, scientists pre-select a relatively small subset of variables (say, twenty) of interest for analysis from a large study with thousands of variables. We consider this pre-selection, with any relationships among them permitted, to be the workload specification; further limiting the queries would problematically bias the data outcomes. We thus assume that relationships between any of the selected variables of interest are permitted. Thus, for the AIM and GEM algorithms, we automatically generate a workload across all variables. 

\paragraph{DP evaluation.} \citet{tao2021benchmarking} evaluated several DP synthesizers' ability to preserve 1- and 2-variable distributions. \citet{jayaraman2019evaluating} studied privacy-utility trade-offs for ML tasks and found that commonly-used $\epsilon$ values and implementations in practice may be ineffective: either unacceptable privacy leakage or unacceptable utility tends to occur.  \citet{hay2016principled} used 1- and 2-dimensional range queries over 27 public datasets to study the influence of dataset scale and shape that had led to inconsistent results in the previous literature. While the datasets are diverse in relevant properties, the tasks are limited, and the link to conclusions drawn is unexplored. \citet{hill2015evaluating} studied the utility of DP on one longitudinal behavioral science dataset involving a sexuality survey motivated by real world attacks based on disclosing pregnancy, finding that theoretical guarantees of DP were generally supported, but that high-dimensional data was a challenge for utility. In this work, we aim to facilitate and standardize these kinds of applied studies.

\paragraph{Reproducibility.}  Numerous reproducibility studies have been attempted in various fields, typically reporting remarkably low rates of success, leading to calls for significant changes to policy and incentive structures underlying scientific funding and publishing~\cite{national2019reproducibility,nosek2018preregistration,munafo2017manifesto}. In the social sciences, \citet{camerer2018evaluating} replicated 21 systematically selected experimental studies published in Nature and Science between 2010 and 2015, finding a significant effect in the same direction as the original study for 13 (62\%) studies, with about half the effect size, on average.  The Reproducibility Project: Cancer Biology~\citet{errington2021reproducibility} attempted to reproduce 193 experiments from 53 papers, but succeeded in only 50 experiments from 23 papers.  They found that only 2\% of papers supplied open data, 0\% of protocols were completely described,  67\% of experiments required modifications to complete, and replication effect sizes were 85\% smaller than in the original findings.   A survey by~\citet{baker20161} found that 52\% of respondents agreed that reproducibility represents a `crisis' for science.   

The terms \emph{reproducibility} and \emph{replicability} are used inconsistently across and even within fields~\cite{barba2018terminologies}. We use reproducibility to mean same data, different code, different team~\cite{barba2018terminologies}.  We use replicability to mean acquiring new data in a new experiment to determine whether the same conclusions hold.  Our methodology amounts to first \emph{reproducing} conclusions found in peer-reviewed papers, and then \emph{replicating} these same conclusions on DP synthetic data. While our focus is neither to reproduce nor to replicate the original study,  our framework supports reproducibility analysis as an intermediate step, and our insights regarding the difficulty (and, often, the impossibility) of reproduction are consistent with prior findings.

\begin{figure}[t!]
\centering
  \includegraphics[width=1.0\columnwidth]{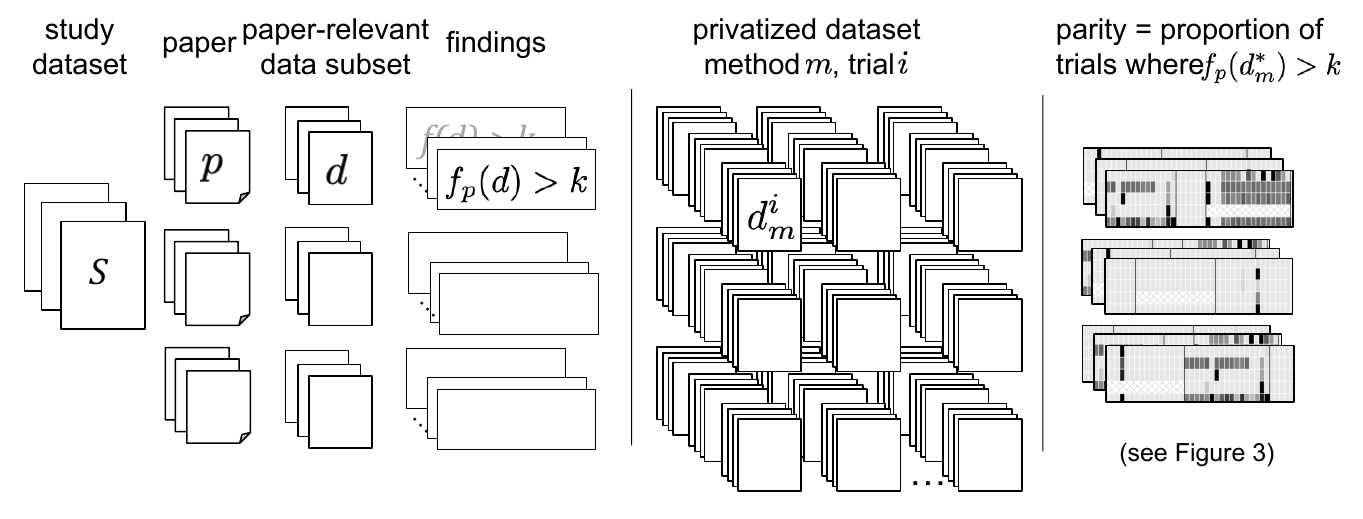}
  \captionof{figure}{Epistemic parity workflow: Each study dataset supports many papers, each using a subset of the features.  The paper's findings are implemented as computable inequalities. We generate many privatized datasets using different random seeds, then compute the proportion of these trials for which the findings hold (Figure \ref{fig:parity}).}
  \label{fig:workflow}
\end{figure}

\section{Preliminaries}
\label{sec:prelims}

Differential privacy (DP) is a strong privacy guarantee that ensures that altering or removing one record from a given dataset does not significantly affect the outcome of an analysis or query. Intuitively, DP prevents an observer of a private output from drawing conclusions about which specific individuals' information was included in the  input. DP is based on the concept of ``neighboring'' datasets, which are datasets that result from the modification or removal of one record that induces a ``neighboring'' dataset. There is a distinction between modification and removal in this definition; establishing which version of the privacy definition is normally done \textit{a priori}. In the scope of the private synthesizers considered by this paper, datasets $X$ and $X'$ are considered neighboring if the removal of a single element $x_i$ from one yields the other (except in the case of PrivBayes; we account for this in our budget allocation).

The conventional understanding of DP is provided in Definition~\ref{def:dp}, while the alternative, zero-concentrated DP ($\rho$-zCDP) is provided in Definition~\ref{def:zcdp}.
\begin{definition}[$(\epsilon,\delta)$-Differential Privacy]
\label{def:dp}
A randomized mechanism ${\mathcal {M}}$ is said to be $(\epsilon,\delta)$-differentially private if, for every pair of neighboring datasets $X$ and $X'$, and all subsets $R$ of possible outputs, the following holds:
$${\displaystyle \Pr[{\mathcal {M}}(X)\in R]\leq e^{\epsilon} \Pr[{\mathcal {M}}(X')\in R] + \delta}$$
\end{definition}
\begin{definition}[Concentrated Differential Privacy ($\rho$-zCDP)\cite{bun2016concentrated}]
\label{def:zcdp}
We use $D_\alpha\left(\mathcal{M}(X)||\mathcal{M}(X')\right)$ to signify
the $\alpha$-R'enyi divergence. Thus, a randomized algorithm $\mathcal{M}$
upholds \emph{$\rho$-zCDP} if, for
every pair of neighboring datasets $X$ and $X'$,
$$D_\alpha\left(\mathcal{M}(X)||\mathcal{M}(X')\right) \leq \rho\alpha, \forall \alpha \in (1,\infty)$$
\end{definition}
These two principles, $(\epsilon,\delta)$-DP and $\rho$-zCDP, are intimately related but scale with different relative privacy parameters. As demonstrated by Bun\etal~\cite{bun2016concentrated}, an established hierarchy of these guarantees is: An $(\epsilon,0)$-DP mechanism gives $\frac{\epsilon^2}{2}$-zCDP, which in turn gives $(\epsilon \sqrt{2 \log(1/\delta)}, \delta)$-DP for every $\delta > 0$.

It is theoretically difficult to compare private mechanisms satisfying differing versions of DP. However, in the case of private synthesizers, practical, utility-based comparisons have become a norm~\cite{rosenblatt2020differentially,tao2021benchmarking,cai2021data,mckenna2022aim}. The methods we consider provide different guarantees: AIM and GEM both give $\rho$-zCDP guarantees, while MST, PATECTGAN and PrivMRF give $(\epsilon,\delta)$-DP guarantees, and PrivBayes gives an $(\epsilon,0)$-DP guarantee. For comparison purposes, all $\epsilon$ parameters are translated using the parameter relationships given above, so as to compare at the same relative privacy settings between synthesizers. As is accepted practice, we set our $\delta$ parameter to be ``cryptographically small,'' at the very most $\frac{1}{n}$ where $n$ is the size of our data, but usually much smaller \cite{dwork2014algorithmic}.

\section{The Epistemic Parity Methodology}
\label{sec:model}

Informally, we say that epistemic parity holds if all conclusions one draws on the \emph{original dataset} also hold on the \emph{synthetic dataset}.  
We define epistemic parity operationally as follows.

\begin{definition}[Epistemic Parity]
Let $D$ be a universe of datasets. A \textit{finding} is a function $D \rightarrow \{0,1\}$.  A \textit{synthesizer} is a function $S : (X, D) \rightarrow D$ where $X\sim U(0,1)$ is a random seed.  Given a dataset $d \in D$, an associated finding $f$, and a DP synthesizer $S$, we say the epistemic parity of $S$ for $f$ on $d$ is the fraction of random trials $i\in X$ such that $f(S(i, D)) = f(D)$.
\end{definition}

The overall workflow to compute epistemic parity is illustrated in Figure \ref{fig:workflow}.  The input is a set of papers, and the output is a set of scores indicating whether findings are supported under various DP synthesizers. 
This framework, and the associated open source software that partially automates it, is intended to expand over time.  Mechanism designers are encouraged to implement new DP methods, and researchers are encouraged to implement new findings from relevant papers.

Beginning from the left in Figure \ref{fig:workflow}: A study is associated with one dataset and potentially many papers.  Each paper typically uses only a subset of the variables in the study.   We assume public access to the data on which the paper's results were computed; our focus is on evaluating DP methods (requiring ground truth) rather than on protecting the privacy of subjects involved in the study.  Any evaluation of DP mechanisms requires access to the original data for interpretation; indeed, inaccessible ground truth undermined the US Census Bureau's efforts to build trust regarding DP~\cite{boyd2022differential}.

Given a paper, we manually identify natural language claims made by the authors as candidates for findings. Though these claims may appear anywhere in the paper, most were found in the results section. Domain expertise of course provides an advantage in this exercise, but we contend that it should always be possible for non-expert readers to identify major claims since the goal of a paper is to communicate findings to a broader audience.

For each claim, we identify the quantitative (or qualitative) evidence that supports the claim, recording the variables involved and methods used.  We then re-implement the analysis to (attempt to) reproduce the salient findings and conclusions in the paper over the original, public dataset, as discussed in Section~\ref{sec:model:repro}. 

While in principle this reproduction step is always possible, in practice, it can be difficult or impossible~\citep{baker20161,national2019reproducibility},and may involve guesswork when the computational details are incomplete.  Moreover, inconsistent reproducibility can introduce bias in our benchmark: we may be more likely to include findings for which computational details are clear, which may be those that are simpler to explain or better-known by the author, as opposed to a representative sample of appropriate methods.  

If the reproduction was successful, we generate $k \times m$ synthetic datasets representing $k$ trials with different random seeds and $m$ different DP methods, and then draw an additional $B$ samples from each seeded DP method.  In our initial benchmark, $k=10$ and $m=5$, and $B=25$. The additional $B$ draws  allow us to bootstrap a confidence interval for each (trained) synthesizer.  That is, there are two sources of randomness: the training procedure used by the mechanism, and the random sampling of the learned model to actually generate synthetic data.  Although each synthetic dataset could be scaled to any number of records --- recall that we are sampling a privatized model --- we always use the same number of records as the original data for each bootstrap sample.  Given this set of synthetic datasets, we again attempt to reproduce the findings using each one.  Finally, we contrast the findings based on original and DP data by measuring the proportion of trials, for each method, where a given finding holds. Our methodology is implemented in an open-source framework, described in Section~\ref{sec:software}.

\subsection{Reproducing Experimental Studies}
\label{sec:model:repro}

We adapt three concepts of reproducibility---\emph{values}, \emph{findings}, and \emph{conclusions}---from \citet{cohen2018three} into a practical taxonomy for reproducing a statistical analysis in a peer-reviewed publication, and implement a software framework that allows us to conduct concrete experiments around this taxonomy (see Section~\ref{sec:software}). 

The atomic element in reproducibility is a \textit{finding}, defined by \citet{cohen2018three} as ``a relationship between the \textit{values} for some reported figure of merit with respect to two or more dependent variables.'' For the purposes of our study, a \textit{finding} consists of a natural language statement (i.e., a \emph{claim}) reported in a publication, along with evidence provided by one or more quantitative or qualitative sub-statements about the analysis.

Evidence for a \textit{finding} consists of a comparison between two or more \textit{values} that can be evaluated as a Boolean condition.  A value may be a scalar
(i.e., $34.1\%$), an aggregated or computed result (i.e., a regression coefficient of $1.2$), or even an implicit threshold expressed in natural language (e.g., ``a low rate'' or ``a strong correlation'').  In these cases, we instantiate the language as a quantitative threshold, applying conventions from the literature when they exist.  For example, a common convention is that Pearson's correlation is considered ``strong'' when $r$ is larger than 0.7.

A special case of a \textit{finding} is a qualitative \textit{visual finding} that often appears in the form of a figure, table or diagram. A figure encodes many potential \textit{findings}; we do not (necessarily) consider each of these sub-findings on their own in our analysis, but rather treat them as a single \emph{visual finding}: we attempt to reproduce the figure itself, and subjectively evaluate its similarity to the original. Consider Figure~\ref{fig:qualitative}, where the top sub-figure shows a percentage  breakdown of drug use across demographic groups from~\citet{fairman2019marijuana}, and the bottom sub-figure shows the same breakdown reproduced on DP synthetic data generated by MST at  $\epsilon=e$. 

Finally, following \citet{cohen2018three}, a \emph{conclusion} is defined as ``a broad induction that is made based on the results of the reported research.'' A conclusion must be explicitly stated in a paper, and comprises one or several \textit{findings}. Conclusions are less concrete than individual findings, and are often at the discretion of the domain expert tasked with interpreting a set of findings. 

\subsection{Generating DP Synthetic Data}
\label{sec:model:dp}

Each of the papers that we reproduced using DP synthetic data derived findings from a subset of the full study's data. For example, HSLS:09 consists of over 7000 columns, but~\citet{jeong2021} used only a subset of 57. We synthesize the subset of data relevant for the reproduced findings and conclusions, as discussed in Section~\ref{sec:model:repro}. In the case where a paper relies on longitudinal data from a study, we collapse the data such that it is ``one row to one person.''

The DP methods for private data release are executed for the range of $\epsilon$ values $\epsilon \in \{ e^{-3}, e^{-2}, e^{-1}, e^0, e^1, e^2\}$, which represents a small to medium privacy regime~\cite{bowen2020comparative}. Each DP mechanism is run 10 times to produce, at each $\epsilon$ value, $10 \times B$ sampled datasets using the same sample size but different random seeds (where $B$ is the bootstrap parameter).  Each DP method involves different hyperparameters and varying levels of tunability, but we use author-recommended settings to avoid biasing results towards our own expertise.  We then re-compute the findings for each sample.

If all findings are reproduced regardless of $epsilon$ or random seed, we say that the DP mechanism achieves complete epistemic parity.  But we measure  parity as the proportion of iterations for which the finding holds.  The goal is to overlook small variations in the exact value in favor of maintaining the relative relationships of the computed statistics for interpretability and practical utility.

\subsection{Analysis of Variability for DP Synthetic Data}
\label{sec:model:variability}
\citet{raisa2022noise} considers a set of rules proposed by Raghunathan and Rubin as a potential approach to calculating uncertainty over the results of DP synthetic data \cite{raghunathan2003multiple}. Consider synthetic datasets $x_1,...x_m \in X$, and test statistic $\tau$. Running $\tau(X_i)$ produces a point estimate $q_i$ with variance $v_i$ for dataset $X_i$. Standard hypothesis testing might calculate a confidence interval and power over the distribution of all $q_i$. In the case where $X$ consists of synthetic data, Rubin's rules for producing such calculations can be applied under normality assumptions \cite{raisa2022noise, raghunathan2003multiple}. In brief, these rules are as follows: to compute confidence intervals, the estimated locations $q_1,...q_m$ and variances $v_1,...v_m$ are combined:
\begin{align}
    \hat{q} = \frac{1}{m} \sum^m_{i=1} q_i \\
    \hat{v} = \frac{1}{m} \sum^m_{i=1} v_i \\
    b = \frac{1}{m-1} \sum^m_{i=1} (q_i - \hat{q})^2
\end{align}

And then we use a $t$-distribution centered at $\hat{q}$, with variance
\begin{align}
    T = \left(1+\frac{1}{m}\right)b-\hat{v},
\end{align}
and degrees of freedom:
\begin{align}
    \text{df} = \left(1-\frac{1}{1+\frac{1}{m}}\frac{\hat{v}}{b}\right)^2(m-1).
\end{align}

This method \textit{crucially} relies on a normality assumption for the distribution over each $\tau(X_i) = q_i$. If the $\tau$ statistic is a mean and variance over a well understood distribution, this could be a reasonable assumption to make. However, in the case of our benchmark, we consider a wide variety of more complex, multivariate statistical tests over multiple high dimensional datasets. For example, weighted least-squares regression produces a set of coefficients for covariates, and we compare the magnitude of these covariates in relation to a target variable of interest. Assuming a normal distribution over covariate values across different randomly initiated trainings of a synthesizer, with the addition of DP noise, is strong. Thus, we are uncomfortable making this normality assumption.

We take a more conservative approach. Instead of computing the statistical power of each point estimate $q_i$, we turn each test into a ``soft'' finding, given informally by Equation~\ref{eq:parity}. The function $finding$ here takes in $\tau$, some ground truth data $q_i^{real}$ and synthetic data $q_i^{synth}$ drawn from a privatized distribution intending to match the distribution of $q_i^{real}$. The parameter $\alpha$ manages the tolerance of the comparison, acknowledging that the DP noise and the nature of synthetic data will certainly alter the true values slightly. $\alpha$ is necessarily assigned according to the context of a $finding$ (for example, if for groups A and B, we compare $\mu_A, \mu_B \in (0,1)$, where in the real data $\mu_A \approx \mu_B$, a tolerance of $\alpha = 0.01$ might suffice).
\begin{align}
\label{eq:parity}
    finding(\tau, q_i^{synth}, q_i^{real}) =  \mathbbm{1}[|\tau(q_i^{synth}) -\tau(q_i^{real})| \leq \alpha]
\end{align}
Recall that the goal of our benchmark is primarily to determine the viability of a synthesizer for reproducing a specific paper's findings, so that we draw the same overall conclusions. Parity is then the proportion of agreement for a given paper and given synthesizer over a fixed set of discrete, statistical tests posed as $findings$. This approach allows us to compare across findings of different types, and to score synthesizers for each paper on the reproducibility of their results.

But how should we address the multiple sources of \textit{variance} in synthetic data? There are two main elements of randomness. Each parametric DP synthesizer is trained conditioned on a random seed, which dictates the privatizing random noise introduced during model fitting. Varying the seed changes the random initialization bias \textit{significantly} for DP synthesizers. Then, after training, once a synthesizer is parameterized, a specific draw from that synthetic distribution is again conditioned on a \textit{different} random seed, which dictates the bias of a particular realized sample from the synthetic distribution. The variance effect for the post training draw is much lower, as no private noise need be added during sampling.

The latter source of randomness (in drawing from a synthetic distribution) is easier to deal with. Conditioned on synthesizer $S_i$, for fixed finding $finding(\tau, q_i^{synth}, q_i^{real})$, we perform a bootstrap over $B$ samples of size $n$ (where $n$ is the size of the real ground truth empirical sample we trained on). Over $B$ datasets, we calculate the average statistic and average variance of $finding$, $\hat{q} = \frac{1}{m} \sum^m q_i$ and $\hat{v} = \frac{1}{m} \sum^m v_i$, as well as a $95\%$ confidence interval.

The former source of randomness (in fitting a synthetic distribution) is more difficult to deal with, as fitting a synthetic distribution is computationally expensive for high dimensional data \cite{rosenblatt2020differentially,tao2021benchmarking,mckenna2022aim}. Despite running extensive tests, we were unable to run enough trainings per synthesizer to say anything convincing about the statistical power of our sample of random synthesizer initializations. Instead, we simply report on the variance of the aforementioned soft finding bootstraps over 10 randomly initialized synthesizers, per paper, per privacy level (around 1400 trained synthesizers in total). Though this is standard practice \cite{bowen2020comparative,bagdasaryan2019differential,tao2021benchmarking}, we believe there's an opportunity to improve the norms here through model sharing and well supported open source synthesizers.
\section{Benchmark Construction}
\label{sec:benchmark}

We used a standardized approach for study and paper selection. Each study, which has an associated dataset, was selected for broad impact (at least 100 papers). For each selected study, we limited our publication search to the past five years, to ensure a focus on modern methods of analysis. Within that five year window, there might still be 10s or 100s of data related publications. We sought papers that meet all of the following criteria:
(1) Are publicly available (so that we could report on their results without violating any permissions); 
(2) Use a publicly available portion of the study dataset (so that we were not trying to replicate analyses conducted on private data using a public subsample); 
(3) Are published in peer-reviewed publications, with a preferences given to high-impact journals; 
(4) Are cited (for papers that are at least two years old); and
(5) Are of a reasonable length (page count $< 30$). 
We describe selected studies and papers below, see our public repository, available at \href{https://github.com/DataResponsibly/SynRD}{https://github.com/DataResponsibly/SynRD}, for details.

\subsection{Selected Studies}
\label{sec:benchmark:studies}
We selected four prominent, federally funded studies, which we now briefly describe. 

\textbf{HSLS:09}: High School Longitudinal Study~\cite{dalton2016high}, is a nationally representative, longitudinal study of U.S. 9th graders who were followed through their secondary and postsecondary years.  We attempted to reproduce four papers that use HSLS:09, and were able to fully or partially reproduce three papers. As HSLS:09 has a single dataset representing the entire period of the study, we did not encounter versioning issues during reproduction.

\textbf{AddHealth}: National Longitudinal Study of Adolescent and Adult Health~\cite{addhealth}, consists of a nationally representative sample of U.S. adolescents in grades 7 through 12 during the 1994-1995 school year.   We attempted to reproduce four papers that use Add Health, and were able to fully or partially reproduce two. The public use data was severely limited in scope (a random subsample of less than 50\% of the original), and we were forced to only consider papers that relied solely on the public use data.

\textbf{NSDUH}: National Survey on Drug Use and Health 2004-2014~\cite{nsfduh}, measures the prevalence and correlates of drug use in the U.S.. We attempted to reproduce four papers that use NSDUH, but we were able to reproduce only one, partially and with substantial effort. The main obstacle was that the study is broken down by year across a fifteen-year time frame and multiple versions of the study for each year have been released. Past years of the study are seemingly updated without a record of what the new version modifies. Thus, we had extreme difficulty finding the breakdown of data version for each year to attempt replication.

\textbf{ACL}: The Americans' Changing Lives Survey~\cite{acls}, is an ongoing longitudinal study of the lives of U.S. adults. The study has several waves, the first of which was conducted in 1986, and each wave follows up with the same sample of U.S. adults, asking a variety of questions to determine how social connections, work, and other factors affect health throughout their lifetimes. We attempted to reproduce 2 papers that use ACL, and were able to partially reproduce both of them. We did not face versioning issues, as the ACL provides a public dataset containing data from each phase and notes where variables have been updated.

\subsection{Selected Papers}
\label{sec:benchmark:papers}

\textbf{\citet{saw2018cross}} use HSLS:09 to examine cross-sectional and longitudinal disparities in STEM career aspirations among high school students. They focus on intersectional interactions between gender, race/ethnicity and socioeconomic status, and assess the progression of ``aspiration'' (\ie stated interest in pursuing a STEM-related career) from 9th to 11th grade of high school.
Methods include singular and trivariate (across race, socioeconomic status and gender)
disparity analysis of quantities of interest, and analysis of disparities among students deemed ``persisters'' (who persist in their STEM interest from 9th to 11th grade) and ``emergers'' (who emerge with STEM interest in 11th grade, having no interest in 9th grade). 
The authors conclude that girls from all ethnic/racial and socioeconomic backgrounds, and lower income Black or Hispanic boys, had substantially lower rates of interest, persistence and development of STEM aspirations. They support their conclusion with a set of findings that mainly catalogue examples of statistically significant gaps in their trivariate demographic analysis of disparities.

This paper was reproducible, with effort. The authors provided an overview of data processing methodology, but failed to specify exact columns and clarify preprocessing procedures (\eg creating ``emerger'' and ``persister'' student sets). However, we were able to reproduce each finding and agree with all conclusions. 

The paper was published in 2018 in \textit{Educational Researcher}, a top education research journal. It has been cited 105 times as of May 2023, according to Google Scholar.

\begin{table*}[]
\centering
\begin{tabular}{lrrrrrrr}
\toprule
 \textbf{Paper}    & \textbf{Sample Size} & \textbf{Variables} & \textbf{Domain Size} & \textbf{Outliers} & \textbf{Mutual Info.} & \textbf{Skewness} & \textbf{Sparsity} \\
\midrule 
\citet{assari2019baseline} & 3361  & 16 & 9.03e+09 & 9 & 0.051 $\pm$ 0.153 & 0.563 $\pm$ 1.557 & 0.253 $\pm$ 0.231 \\ 
\citet{fairman2019marijuana}            & \textbf{293581} & 6 & 2.03e+05 & 0 & 0.255 $\pm$ 0.432 & 0.185 $\pm$ 0.462 & 0.174 $\pm$ 0.165 \\
\citet{fruiht2018naturally}            & 4173 & 11 & 2.20e+05 & 6 & 0.104 $\pm$ 0.256 & 0.607 $\pm$ 1.694 & 0.394 $\pm$ 0.183 \\ 
\citet{iverson2021high}         & 1762 & 27 & 5.71e+15 & 5 & \textbf{0.004 $\pm$ 0.010} & NaN & 0.307 $\pm$ 0.180 \\ 
\citet{jeong2021} & 15054 & 57 & \textbf{7.04e+42} & 32 & 0.020 $\pm$ 0.026 & 0.338 $\pm$ 2.850 & 0.261 $\pm$ 0.166 \\ 
\citet{lee2021ability} & 14575 & 9 & 5.11e+17 & 5 & 2.862 $\pm$ 1.242 & 0.080 $\pm$ 0.440 & 0.111 $\pm$ 0.156 \\ 
\citet{PierceQuiroz2019} & 1585  & 17 & 7.19e+11 & 11 & 0.030 $\pm$ 0.050 & 0.001 $\pm$ 1.050 & 0.146 $\pm$ 0.158 \\ 
\citet{saw2018cross} & 20242 & 9 & 4.30e+04 & 3 & 0.143 $\pm$ 0.145 & 1.291 $\pm$ 1.218 & 0.354 $\pm$ 0.171 \\
\midrule 
Adult~\cite{adult} & 32561 & 15 & 9.06e+14 & 96 & 0.066 $\pm$ 0.053 & 17.455 $\pm$ 22.992 & 0.125 $\pm$ 0.164 \\ 
Mushroom~\cite{mushroom} & 8124 & 23 & 2.44e+14 & 74 & 0.199 $\pm$ 0.209 & 6.211 $\pm$ 8.955 & 0.297 $\pm$ 0.219 \\ 
\bottomrule
\end{tabular}
\caption{Properties and meta-features of the datasets in our benchmark, and of two datasets that are commonly used for DP benchmarking, Adult and Mushroom. Mutual Information, Skewness and Sparsity are the \textit{average} for each of these metrics across all variables in the dataset. Our results (Section \ref{sec:experiments:quant}) reinforce that synthesizers may struggle with large sample sizes (\citet{fairman2019marijuana}), large domain sizes (\citet{jeong2021}), and low mutual information (\citet{iverson2021high}).} 
\label{tab:complexity}
\end{table*}

\textbf{\citet{lee2021ability}} use HSLS:09 to identify factors that affect the performance of students on the 11th grade math exam. The authors examine ``low teacher support'' as an adverse factor, and self perceptions of math ``ability'' and ``parental support'' as protective factors.
Factors are constructed by aggregating across relevant survey responses via a weighted average. The authors control for demographic variables and for historical math performance to isolate the effects. Pearson correlation is computed across these variables and demographic information. Linear regression models are trained to predict math performance, with different interactions between variables.   
The authors conclude that high ``ability self-concept'' and ``parental support'' can make an adolescents more resilient to poor teaching.

This paper was partially reproducible, with substantial effort. The authors detailed their aggregation techniques reasonably well, and also included a helpful table for survey questions in the appendix. They did not detail their regression techniques, although results of a simple weighted linear regression turned out to align well with their findings. We were unable to reproduce a finding that involves a covariance slope analysis figure from a complex R package. 

The paper was published in 2021 in \textit{Journal of Adolescence}, a reputable adolescent psychology research journal, with 8 citations as of May 2023, per Google Scholar.

\textbf{\citet{jeong2021}} use HSLS:09 to interrogate potential racial bias in classifiers that predict student performance on a standardized 9th grade math exam.
The authors assign each student to one of two racial groups --- White/Asian (privileged) or Black/Hispanic/Native American (disadvantaged).  They trained random forest, SVC and logistic regression models to predict whether a student would receive a top-50\% or a bottom-50\% test score, and measured accuracy, FPR, FNR, and predicted base rate, both overall and for each racial group.
They find that FPR was almost twice as high for the privileged students, while the FNR was twice as high for the disadvantaged.  The authors concluded that privileged students were given the benefit of the doubt, while the disadvantaged were systematically under-estimated by the classifiers.

We found this paper to be reproducible, with effort.  The authors did not specify how the data was preprocessed (\eg how missing values were imputed) or how it was split into training and test.  We were ultimately able to reproduce the results sufficiently to agree with the conclusions, but were unable to reproduce the values in the findings exactly.

This paper was published in 2021 in the \emph{NeurIPS Workshop on Math AI for Education}, with 1 citation as of May 2023, according to Google Scholar.

\textbf{\citet{fruiht2018naturally}} use AddHealth to investigate the role of naturally occurring mentors in the educational outcomes of first-generation college students. The information collected included demographic information, parental educational attainment, naturally occurring mentors, the nature of support by the mentor in question, followed by the educational attainment of the survey participant as an adult. 
The authors fit a statistical mediation model by \citet{preacher2008asymptotic} (PROCESS Model, variation 1) to test direct and interaction effects of parental educational attainment and mentorship on students' educational attainment.  
They conclude that having at least one parent who graduated from college, or having a mentor, was positively associated with higher educational attainment, and, further, that African Americans experienced lower educational attainment than other participants. They further conclude that having a mentor moderated the relationship between participants' educational outcomes and those of their parents. First-generation college status, the presence of a mentor, and race (particularly for students identifying as African Americans) all had a direct effects on educational attainment as an adult. 

This paper was partially reproducible, with substantial effort. Beyond the AddHealth study, the authors reported on findings made though manual qualitative coding of free text responses regarding the nature of support provided by the mentor, but did not make the coding scheme publicly available, and so we were unable to reproduce findings reliant on the scheme.  For other findings, lack of detail about pre-processing, such as how individuals with multiple races were categorized into racial groups, prevented a perfect replication of many of the values reported in the paper, although we were able to reproduce the observed trends and agree with the authors conclusions.   

The paper was published in 2018 in the \textit{American Journal of Community Psychology}, a reputable journal covering community health. It has been cited 57 times as of May 2023, according to Google Scholar.

\textbf{\citet{iverson2021high}} analyze the effect of having played high school football on depressive and suicidal tendencies in men later in life using the first wave (1994-95) and the most recent wave (2016-2018) of AddHealth. The authors conduct a bivariate analysis across two groups of men (those who did or did not play football in high school), controlling for demographics and risk factors, and report on simple percent comparisons, statistical significance and odds ratios.
The authors did not find a direct effect of playing football in high school on depressive tendencies, but did find that those who had depressive tendencies in adolescence were more likely to keep those tendencies later in life. 

This paper was easy to reproduce. The authors provided precise guidance on which questions (columns) they used and applied relatively simple methods. We were able to exactly match nearly all values reported in the paper, reproducing all findings.

The paper was published in 2022 in \textit{Frontiers of Neurology}, a top journal in the field, with 2 citation as of May 2023, according to Google Scholar.

\textbf{\citet{fairman2019marijuana}} use NSDUH to investigate predictors and potential consequences of initiating the use of marijuana before other types of substance (\eg cigarettes and alcohol) for U.S. youth.  The primary methods of analysis were counts and percentage comparisons by group, and computing adjusted relative risk ratio (aRRR) and adjusted odds ratio (aOR).
The authors found that using marijuana first was predictive of future heavy substance use, and showed disparate demographic impact on minority groups of this trend. 
They also analyzed associations between using marijuana first and demographic group membership and found that those using marijuana first were more likely to be male (vs. female), older (vs. younger), and Black, American Indian/Alaskan Native, multiracial, or Hispanic (vs. White or Asian).  

This paper was partially reproducible, with substantial effort. The main obstacle to reproducibility is the inappropriate versioning of the data by NSDUH, as discussed in Section~\ref{sec:benchmark:studies}.  Additionally, the paper inadequately describes their calculations of aRRR and aOR, making them difficult to replicate.  Ultimately, while we were unable to reproduce exact reported values, we did reproduce general trends and agree with most conclusions drawn in the paper.

The paper was published in 2019 in \emph{Prevention Science}, the official journal of the Society for Prevention Research, and has 48 citations as of May 2023, according to Google Scholar.

\textbf{\citet{assari2019baseline}} use ACL to investigate the predictive role of obesity on long-term risk of mortality due to cerebrovascular disease, and, further, to test racial variation in this effect. The authors use the first wave of ACL (1986) to establish baseline factors, including obesity as the main predictor, and demographics, socioeconomic status, health behaviors, and health status as baseline covariates.  They used data from subsequent waves to determine time to death due to cerebrovascular causes.  In the first wave of ACL, race was modeled as a binary variable, with participants identifying either as non-Hispanic White or as non-Hispanic Black. 
The authors performed univariate, bivariate, and multivariable data analysis and concluded that baseline obesity did not predict cardiovascular death in the overall sample.  They then ran race-specific analysis and found that obesity is predictive of cerebrovascular death for Blacks but not Whites.

This paper was partially reproducible, with substantial effort. The authors did not state exactly which variable was used to produce the different features of their analysis, and, more specifically, they did not state how they identified individuals for whom death was caused by cerebrovascular disease. To reverse-engineer their analysis, we used history of cerebrovascular disease combined with death to represent this feature. This allowed us to successfully reproduce 16 of the 18 findings we attempted. 

This paper was published in 2019 in \emph{The International Journal of Environmental Research and Public Health}, and has 10 citations as of May 2023, according to Google Scholar.

\textbf{\citet{PierceQuiroz2019}} use ACL to study how social support and social strain stemming from spouses, children, and friends impact emotional states.
The authors pre-processed ACL data to include a subset of individuals whose responses were present in all waves they analyzed, and each of whom had a spouse, a child, and friends when responding to the survey.  They modeled the dependent variables, namely, positive (resp. negative) emotional state, by combining participants' responses to two (resp. three) survey questions.  They modeled independent variables, namely, support or strain from spouses, children, and friends, by combining answers to two survey questions for each.  Finally, they controlled for family income, education, age, and ``number of confidants'' (number of people with whom the respondent shares their feelings or concerns).
The authors used a mixed-effects model, and found that spousal support increases positive emotions, while spousal and child-based strain increases negative emotions over time.  

This paper was partially reproducible, with substantial effort. The authors did not clearly state which variables they used as basis for demographic attributes (for which they controlled), and we were thus unable to fully reproduce their analysis. We were able to reproduce 10 out of 14 findings stated by the authors.

This paper was published in 2019 in \emph{The Journal of Social and Personal Relationships}, and has 26 citations as of May 2023, according to Google Scholar.

\paragraph{Note on study/dataset dimensionality.}  We did not explicitly filter papers based on the of size dataset utilized for their results. Each of the studies we considered was extremely high dimensional (many thousands of variables), but the corresponding papers in our benchmark each follow a standard subsetting procedure, where they select a small collection of variables of interest and evaluate on those. Thus, our benchmark datasets are not as high-dimensional as some other benchmark challenge sets \cite{mckenna2022aim}.

\subsection{Comparison to Other DP Benchmarks}
\label{sec:benchmark:comparison}

Our benchmark papers introduce a collection of 8 new datasets. In this section, we briefly analyze these new datasets, and discuss characteristics that differentiate them from standard ML datasets, like Adult~\cite{adult} or Mushroom~\cite{mushroom}, which were used in prior DP benchmarks~\citep{hay2016principled, rosenblatt2020differentially}. We perform this analysis through the lens of principled dataset and ML task characterization, sometimes called ``meta-learning.'' Prior work in the meta-learning community provides guidance~\cite{pinto2016towards,Rivolli2018CharacterizingCD} and tooling~\citep{pymfe2020}, of which we take advantage. 

In Table \ref{tab:complexity}, we show several properties and meta-features for eight datasets from our benchmark, as well as for two popular datasets from the UCI Machine Learning repository~\citep{dua2019UCI}, Adult~\cite{adult} and Mushroom~\cite{mushroom}. We highlight the following four meta-features that we suggest could be relevant to DP data synthesis:

\textbf{Number of outliers} is calculated as the number of values that fall outside of the range $[\bar{x} - 1.5 * IQR, \bar{x} + 1.5 * IQR]$,  summed across all numerical variables. Outliers present a challenge to data privatization, as they are easily identifiable.

\textbf{Mutual information (mean, standard deviation)} is calculated for each pair of features. If the joint distribution of two features in the sample $X, Y$ is $P_{(X,Y)}$, and the marginal distributions are $P_X$ and $P_Y$, the mutual information is defined as: 
$$I(X;Y) = D_{\mathrm{KL}}( P_{(X,Y)} \| P_{X} \otimes P_{Y} )$$ 

Here, $D_{\mathrm{KL}}$ is the Kullback–Leibler divergence. DP synthetic data algorithms like PrivMRF, MST, PrivBayes and AIM are, at their core, interested in \textit{preserving} mutual information between features, but this is challenging given the constrained nature of model fitting (often relaying on a small set of 2- or 3-way marginal queries) and the addition of noise for privatization. 

\textbf{Skewness (mean, standard deviation)} of a sample is calculated according to the formula for adjusted Fisher-Pearson standardized moment coefficient:

$$\left( \frac{m-1}{n} \right)^{3/2} \frac{m_3}{m_2^{3/2}} \text{ where } m_i=\frac{1}{N}\sum_{n=1}^N(x[n]-\bar{x})^i$$ 

This is an unbiased estimate that gives similar results to other popular skewness measures for large samples, but can vary for smaller and moderate-sized samples~\citep{joanes1998ComparingMO}. The regularity of variables in a dataset (the level of asymmetry in the underlying distributions) affects their ease of replication.

\textbf{Sparsity (mean, standard deviation)} is defined as a normalized ratio of the number of samples over the number of unique values: $$\frac{1}{n-1} \frac{n}{\phi_v - 1}$$ where $\phi_v$ is the number of distinct values in our sample. Sparser data may be harder to capture through noisy marginal measurements.

Table~\ref{tab:complexity} demonstrates that the datasets in our benchmark are diverse and cover a wide range of values for the aforementioned metrics. Interestingly, one of our most challenging datasets to reproduce, \citet{iverson2021high}, had the lowest average mutual information score and one of the highest sparsity scores. A majority of the synthesizers we test depend on mutual information to select the marginal measurements for distribution learning. Selecting the most relevant 2-way marginals when mutual information is uniformly \textit{low} and there are many features is clearly a challenge. We further note that Adult, a common challenge dataset, had uniquely skewed distributions, which aligns with prior work suggesting that this dataset is unrepresentative/idiosyncratic, and so one should hesitate when using it for evaluation and benchmarking~\cite{ding2021retiring}. 
\section{Extensible Benchmarking Package}
\label{sec:software}

We organize the epistemic parity workflow and existing papers and datasets into an open-source differentially private (DP) synthetic data benchmarking package called \synrd, available at \href{https://github.com/DataResponsibly/SynRD}{\url{https://github.com/DataResponsibly/SynRD}}.  This package helps answer the question: ``Can a DP synthesizer produce private (tabular) data that preserves scientific findings?'' In other words, to what degree does a DP synthesizer satisfy epistemic parity for a particular set of findings?

We used \synrd to automate empirical evaluation of epistemic parity of five state-of-the-art DP  synthesizers (see Section~\ref{sec:related}) over all publications in our benchmark (see Section~\ref{sec:benchmark}).  The \synrd package is built to be naturally extensible to include additional DP synthesizers, datasets, publications, and finding types. 

The main classes implemented by \synrd are \texttt{Synthesizer}, \texttt{Publication}, \texttt{Finding}, and \texttt{Benchmark}.  
The \texttt{Synthesizer} class provides a unified interface to the implementations of five DP synthesizers, specifying recommended parameter values for each, and implementing the \texttt{fit} and \texttt{sample} methods.  This class wraps implementations of MST, PATECTGAN and AIM from the \href{https://github.com/opendp/smartnoise-sdk}{SmartNoise} package, and an implementation of PrivBayes from the \href{https://github.com/DataResponsibly/DataSynthesizer}{DataSynthesizer} package~\cite{DBLP:conf/ssdbm/PingSH17}. PrivMRF is implemented separately.

The \texttt{Publication} class contains descriptive attributes of each paper, specifies input files and any necessary data transformations.  This class also contains a collection of findings, described next. The \texttt{Finding} class implements each analysis method from Table~\ref{tab:methods} as finding type, and automates the process of checking whether a finding is reproduced over a given dataset.  

Finally, the \texttt{Benchmark} class takes as input a publication, along with execution parameters (such as the number of bootstrap samples), checks all findings for that publication over the real data, generates synthetic datasets for each DP synthesizer, checks findings over synthetic data, and, finally, generates an epistemic parity score for each (synthesizer, finding) pair, and for the synthesizer overall (over all findings). 

An example instance of \texttt{Publication}, for~\citet{saw2018cross}, along with an implementation of evaluating epistemic parity for MST using \texttt{Benchmark}, is shown below.  

\begin{python}[language=Python]
from SynRD.papers import Saw2018Cross
from SynRD.benchmark import Benchmark
from SynRD.synthesizers import MSTSynthesizer
benchmark = Benchmark()
B = 25 # Bootstrap parameter
synth = MSTSynthesizer(epsilon=1.0)
papers = benchmark.initialize_papers([Saw2018Cross])
for paper in papers:
    synth.fit(paper.real_dataframe)
    dataset = synth.sample(len(paper.real_dataframe) * B)
    paper.set_synthetic_dataframe(dataset)
    benchmark.eval(paper, B=B)
\end{python}

An example function that implements checking a finding that ``Obesity at baseline was not associated with cerebrovascular death in the pooled sample'' from~\citet{assari2019baseline}, is shown below.

\begin{python}[language=Python]
def finding_6_9(self):
corr_df = self.get_corr()
corr_obesity_death = \
    corr_df['Obesity'].loc['Cerebrovascular death']
soft_finding = abs(corr_obesity_death) < 0.05
\end{python}

To support empirical evaluation of our benchmark, we pre-processed publication-relevant subsets of the ICPSR studies, and deposited them in the Harvard Dataverse repository.  The auxiliary \texttt{DatasetLoader} class retrieves this data, and makes it available to an instance of the \texttt{Publication} class.

\section{Results}
\label{sec:results}

\begin{figure*}[htp]
\label{fig:main}
 \includegraphics[clip,width=\textwidth]{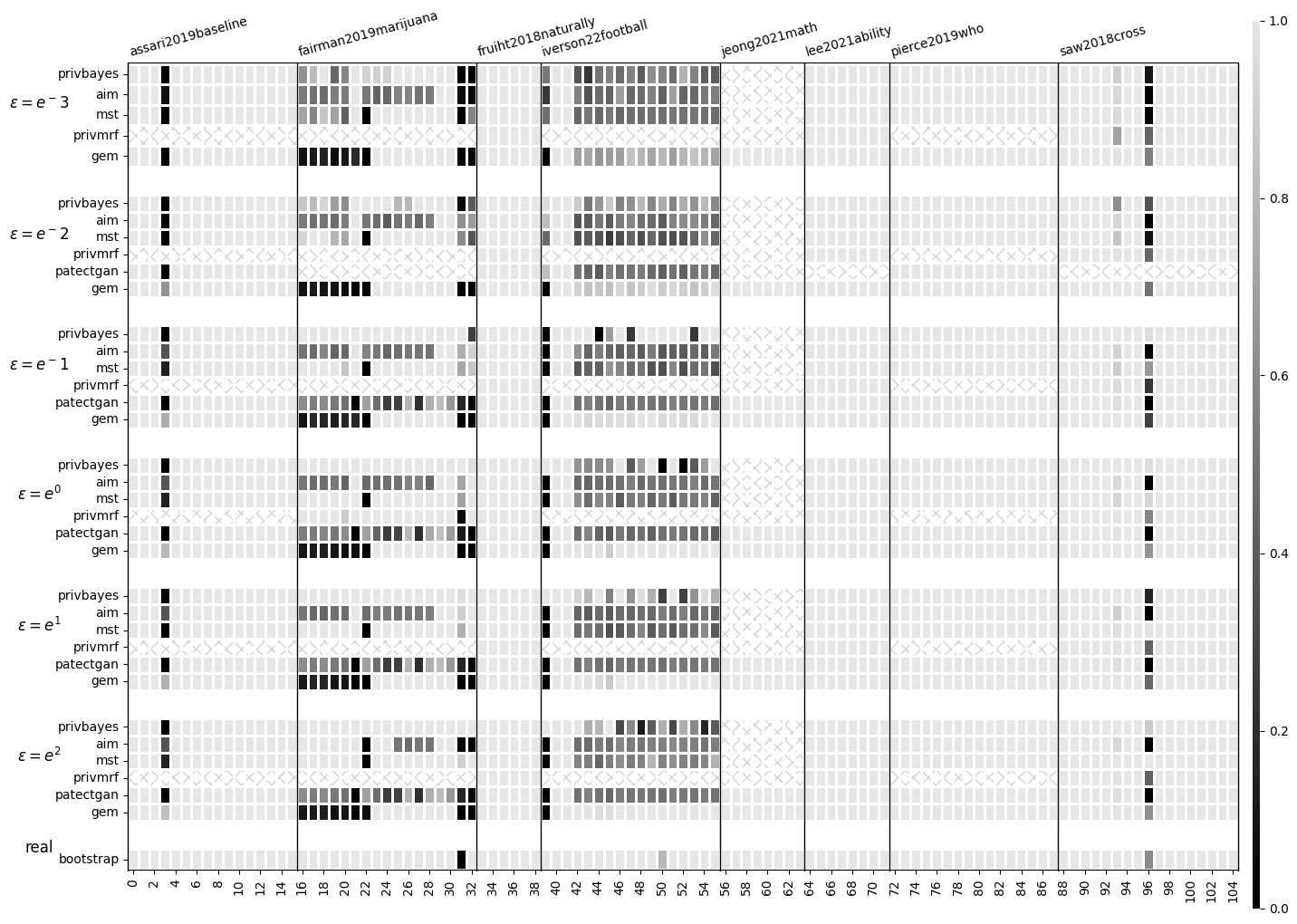}%
\caption{Epistemic parity for six competitive mechanisms for synthesizing data across four $\epsilon$ values ($e^{-3}$,$e^{-2}$,$e^{-1}$, $e^0$, $e^1$, $e^2$).  All mechanisms achieve perfect parity on  \citet{fruiht2018naturally} and \citet{lee2021ability}, and all but one achieve perfect parity on \citet{PierceQuiroz2019}.  Only PATECTGAN can scale to support~\citet{jeong2021}.  All methods struggled with the high dimensionality of \citet{iverson2021high}. PrivMRF was too slow to be viable; we report results only for $\epsilon=e^0$.  Only PrivBayes and MST achieved reasonable parity for \citet{fairman2019marijuana}.  For datasets associated with \citet{assari2019baseline} and ~\citet{saw2018cross}, only one finding was difficult to reproduce, and all methods struggled.  Surprisingly, parity is relatively insensitive to $\epsilon$.}
\label{fig:parity}
\end{figure*}

\paragraph{Computational resources.} Our benchmark consists of eight papers, each evaluated on six synthetic data algorithms for six values of $\epsilon$, for a total of 36 mechanisms for each paper, each repeated with 10 random seeds. We draw 25 samples of size $n$, where $n$ is the real data sample size, and bootstrap over this set of samples when calculating average parity over our finding set. Benchmarking extensively with DP synthesizers is computationally expensive \cite{mckenna2022aim,rosenblatt2020differentially,tao2021benchmarking}. Fitting many synthesizers ($\sim$8 paper datasets $\times$ 10 trainings $\times$ 6 $\epsilon$-privacy regiments) 
took 100s of compute hours. Training PrivMRF and PATECTGAN was done using NYU's Greene High Performance Computing cluster using A100 and RTX8000 NVIDIA GPUs with 80GB and 48GB of RAM respectively. CPUs from that same cluster were used to train AIM, MST, PrivBayes, and GEM. The benchmark itself (assessing parity per paper) was also run on the cluster. Though this was computationally less intense (with some exceptions), bootstrapping to report variance per synthesizer did incur additional overhead.

\subsection{Quantitative results}
\label{sec:experiments:quant}
Figure~\ref{fig:parity} shows parity for all findings across all papers, for each of the five synthesizers, with $\epsilon$ regimens of $e^{-3}$, $e^{-2}$, $e^{-1}$, $e^{0}$, $e^{1}$, and $e^{2}$. Darker color indicates lower average parity, while lighter indicates higher average parity. Each paper is a block of rectangles, where the $x$-axis represents \emph{findings} and the $y$-axis shows the five synthesizers. The crosshatched cells indicate that a synthesizer was unable to fit to a dataset in under 6 hours. 

\paragraph{Baseline: Bayesian bootstrap.}
An important element of our evaluation framework is an accurate assessment of replicability and variability. We subject private synthesizers to a rigorous bootstrap process to assess the epistemic parity of findings presented in each paper; however, it is then reasonable to ask whether the findings themselves, assessed via a bootstrap procedure on the \textit{real} data, would hold up under similar scrutiny? In order to check this, we perform a standard Bayesian bootstrap procedure \cite{rubin1981bayesian}. The Bayesian bootstrap assigns weights to each observation in a given dataset, where those weights $\textbf{w} = (w_1, w_2, ..., w_n)$ are drawn from a symmetric Dirichlet distribution, i.e. $\textbf{w} \sim \mathcal{D}(1, 1, ..., 1)$ (the Dirichlet distribution is a standard Bayesian prior, a smoothed version of the multinomial distribution). We then compute the epistemic parity of each finding with these new weighted samples for $B$ iterations of the bootstrap, matching the number of  iterations of the bootstrapping procedure on the private synthetic data. Each statistic is then an independent and identically distributed realization from the posterior predictive distribution, which offers a Bayesian perspective on the data-generating mechanism. The variability among these samples reflects the uncertainty about the underlying population.

The final row, labeled ``real, bootstrap,'' in Figure~\ref{fig:parity} shows the results of our Bayesian bootstrapping control procedure. We note that over $97\%$ of our findings are reproduced in $100\%$ of our Bayesian bootstrap iterations. For the remaining inconsistent three findings over the bootstrap, it is unfair to expect the private synthesizers to have higher epistemic parity than the bootstrap control.

\paragraph{Epistemic parity: overall performance.} The overall performance of the synthesizers was impressive. All six synthesizers achieved $100\%$ parity for~\citet{lee2021ability}, and~\citet{fruiht2018naturally}. Besides PrivMRF (which was computationally infeasible to fit to the data), AIM, MST, PrivBayes, PATECTGAN, and GEM achieved $100\%$ parity for~\citet{PierceQuiroz2019} as well.
Both~\citet{saw2018cross}, and~\citet{assari2019baseline} also had very high levels of parity between findings on real and on synthetic data, although each of these papers had one finding that was difficult to reproduce  for any synthesizer in any privacy regime.

Two of the papers provided the greatest challenge, and the most interesting results, across privacy regiments and synthesizer types: \citet{fairman2019marijuana}, and \citet{iverson2021high}. These papers were challenging for very different reasons. \citet{fairman2019marijuana} had the second-smallest domain size, and the fewest variables. However, it had by far the largest sample, consisting of nearly 300K records. This combination made it very sensitive to noise in marginal measurements (as they are essentially counts), in turn making the findings difficult to replicate in low-privacy settings. Still, PrivBayes and MST exhibited impressive performance in comparison to AIM, PATECTGAN and GEM. On the other hand, \citet{iverson2021high}, had both one of the largest domains and the most variables of the papers in our benchmark, as well as a low mutual information between variables. No synthesizer, with the exception of GEM, exhibited convincing parity performance on this paper.

\begin{figure}[t!]
\centering
\includegraphics[width=1.05\columnwidth]{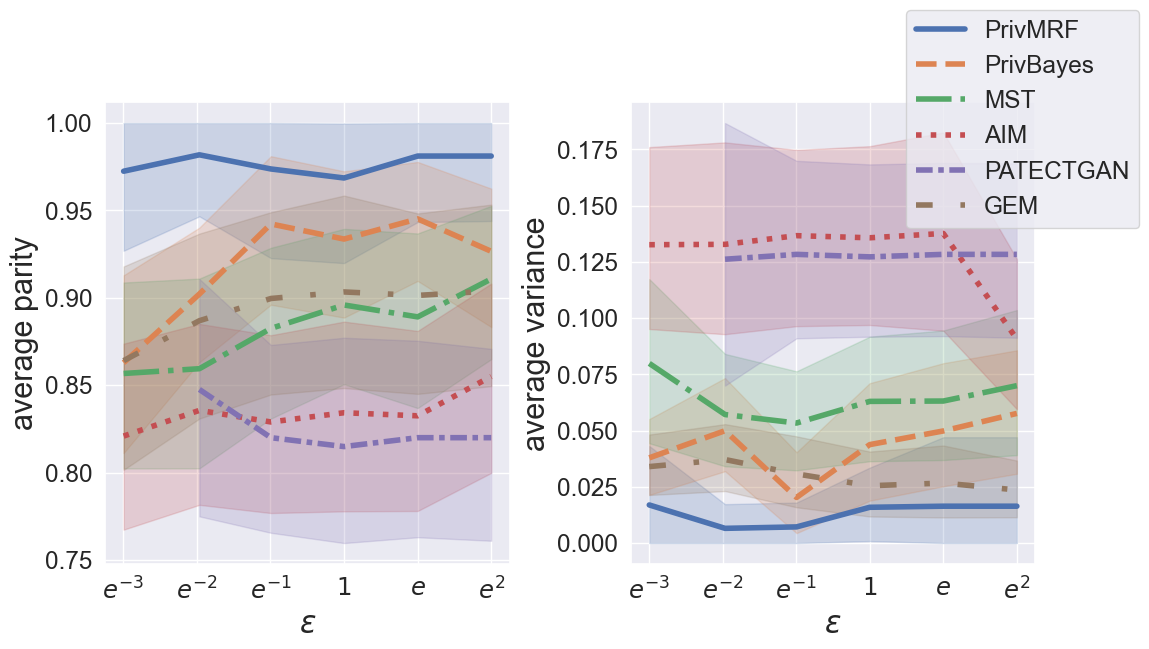}
\caption{Average epistemic parity across papers achieved by AIM, PrivMRF, MST, PrivBayes, PATECTGAN, and GEM as a function of the privacy parameter 
$\epsilon \in \{ e^{-3}, e^{-2}, e^{-1}, e^0, e^1, e^2 \}$.  Parity, on the $y$-axis, ranges between 0 and 1, and represents the fraction of reproduced findings over 10 random seedings of each synthesizer, bootstrapped over 25 random draws from each seed, for 250 total synthetic data draws per paper at each $\epsilon$.}
\label{fig:combinedlines}
\end{figure}

GEM was the strongest performing synthesizer on one paper (\citet{iverson2021high}).  For the other papers in our benchmark, neither PATECTGAN nor GEM were the strongest performing. However, these methods were the most computationally tractable on high-dimensional large-domain data, and were the only methods that were feasible to run on~\citet{jeong2021}, where they both achieved $100\%$ parity. 

Interestingly, PrivBayes often outperformed MST on our benchmark. We believe that this can be explained by two factors: (1) MST is tailored to work on high-dimensional datasets such as NIST, where explicitly parameterizing a conditional structure (like PrivBayes does) is costly and unstable, while the datasets in our benchmark are relatively low-dimensional; and (2) the findings that comprise the epistemic parity metric are based on conclusions that often rely on conditional relationships, which PrivBayes represents explicitly, while MST does not.
 
PrivMRF was the slowest synthesizer to run, and required a GPU.  This requirement limited our ability to fully assess the capabilities of PrivMRF, although we observe that it performed well on the datasets on which it was able to run successfully.  PrivBayes was the second-slowest method to run, due to a known limitation in handling high-dimensional data.  However, PrivBayes performed competitively on datasets on which it was able to run successfully.
    
Notably, no synthesizer succeeded across all papers, and, remarkably, some findings were \textit{never} reproduced by any of the synthesizers. We will revisit these findings further in Section~\ref{sec:experiments:qual}.

\paragraph{Epistemic parity across $\epsilon$ values.} Figure~\ref{fig:combinedlines} compares synthesizer performance across reasonable $\epsilon$ values, shown on the $x$-axis in both sub-figures. The left side of the figure shows aggregated epistemic parity as the percentage of reproduced findings on the $y$-axis, over all iterations of each synthesizer, averaged over all publications in our benchmark.   We observe that synthesizer performance (average parity) improves --- although not substantially --- for higher $\epsilon$ values for marginals-based methods PrivMRF, MST, and AIM.   At the smallest values ($\epsilon = e^{-3}, e^{-2}$), the performance of PrivBayes, AIM, and MST all begin to noticeably (and understandably!) degrade, especially on certain findings (e.g., 16-21).  Interestingly, PrivBayes achieves best performance at $\epsilon=e$, and PATECTGAN and GEM appear insensitive to the value of $\epsilon$.   These trends are consistent with the observations in Figure~\ref{fig:parity}, and support the choice of $\epsilon=e$ as a reasonable privacy budget.    Overall, we observe that restricting the privacy budget to $\epsilon = e^{-3}$) does not significantly affect the ability of the synthesizers to reproduce the ``easy'' findings, while increasing it to $\epsilon = e^{2}$ does not help with reproducing the ``difficult'' findings.  We conjecture that the modeling structure employed by the synthesizer is more important than the scale of private noise. 

The right side of Figure~\ref{fig:combinedlines} shows average variance of epistemic parity.  We observe that variance is lowest for PrivMRF, followed by PrivBayes.  Further, we observe that the value of $\epsilon$ has little impact on parity variance; AIM is the only synthesizer that benefits from a higher value of $\epsilon$ in terms of reduced average parity variance.

The observation that epistemic parity is insensitive to $\epsilon$ is significant.  It suggest that our metric is substantially different compared to other metrics that were previously used for assessment of DP synthesizers.  Parity may provide insight into a more fundamental question about whether a DP synthesizer's \textit{methodology} --- the types of measurements it takes to constitute a synthetic distribution --- is appropriate to preserve the statistical properties of the dataset that are necessary to reproduce \textit{findings}. 

\subsection{Qualitative results}
\label{sec:experiments:qual}

\begin{table}[t!]
\begin{center}
\small 
\begin{tabular}{rlc}
\hline
& Descriptive Statistics & 8\\
    \cline{2-3}
\multirow{2}{*}{Regression}
    &Between-Coefficients & 4\\
    &Fixed Coefficient (Sign) & 2\\
    \cline{2-3}
\multirow{3}{*}{Causal Paths}
    &Variability & 1\\
    &Interaction & 1\\
    &Coefficient Difference & 19\\   
    \cline{2-3}
\multirow{4}{*}{Logistic Regression}
& PBR & 2\\
& FNR & 2\\
& FPR & 2\\
& Accuracy & 2\\
    \cline{2-3}
\multirow{2}{*}{Mean Difference}
    &Between-Class & 24\\
    &Temporal (FC) & 26\\
    \cline{2-3}
\multirow{2}{*}{Correlation}
    & Pearson & 12\\
    & Spearman & 1\\
    \hline
\end{tabular}
\end{center}
\caption{Methods used in benchmark papers, each corresponding to a type of \emph{finding} in our framework.}
\label{tab:methods}
\end{table}

\paragraph{Epistemic parity across finding types.}  Table~\ref{tab:methods} summarizes the methods used in the publications in our benchmark, each corresponding to a type of finding.   We observe \emph{Mean Difference} (both \emph{Between-class} and \emph{Temporal}) is by far the  most common finding type, followed by \emph{Coefficient Difference}.  Whether a finding can be reproduced over DP synthetic data depends on several factors, including dataset size (as in~\citet{fairman2019marijuana}) and dimensionality (as in~\citet{iverson2021high}).  However, finding type likely also plays a role: The majority (19 out of 26) of \emph{Mean Difference / Temporal} findings are in these two papers that were difficult to reproduce.  However, we must be cautious to interpret this as a trend: the remaining 7 findings of type \emph{Mean Difference / Temporal (FC)}  were in~\citet{saw2018cross}, and they were reproduced successfully by all synthesizers.  In what follows, we qualitatively evaluate the impact of finding type (and, possibly, of other properties of the finding) on its reproducibility over DP synthetic data.

That some findings are easier to reproduce than others is unsurprising. Though each synthesizer relies on a fundamentally different approach to replicating the joint distribution across all of the data, they each struggle with high dimensional data.  Further, for general-purpose synthetic data, PrivMRF, AIM, MST and PrivBayes prioritize lower dimensional 2- or 3-way relationships among variables, and thus it is unsurprising that simple mean comparison findings, and even some bi-variate correlations, are easily preserved by these methods.

We were surprised at the high number of findings across all our papers (even those that we were unable to replicate) relying only on 1- or 2-dimensional comparisons: The low-dimensionality suggests that earlier empirical studies (including~\citet{tao2021benchmarking} and~\citet{hay2016principled}) may be suitable as proxy tasks. Targeted improvements to the synthesizers may allow us to simultaneously support high utility for individual findings and their composition into broad conclusions.

Next, we consider three findings that were difficult regardless of synthesizer or privacy regimen:  \#4 (\citet{assari2019baseline}),  \#39 (\citet{iverson2021high}), and \#96 (\citet{saw2018cross}), see Figure~\ref{fig:parity}.

Finding \#4 is of type \emph{Descriptive Statistics}.  It is based on the text statement ``Similarly, overall, people had 12.53 years of schooling at baseline (95\%CI = 12.34-2.73).'' Finding \#39 is also of type \emph{Descriptive Statistics}, and is based on a somewhat longer text statement that refers to specific percentages of individuals being diagnosed with specific disorders (5 such pairs of statistics in total). Finding \#94 is of type \emph{Mean Difference / Between-class}.  It's based on the text statement ``From a longitudinal perspective, students from the two lower SES groups—low-middle and low SES groups—had significantly fewer persisters (31.9\% and 29.9\%) and emergers (6.1\% and 5.4\%) than their high SES peers (45.1\% and 9.0\%, respectively).'' 

These findings were difficult to reproduce because they give specific measurements for variables with large domains. Larger domains require proportionally more DP noise to privatize, and so the learned distribution over each of these variables was too noisy to reproduce these findings within our specified tolerance.  

\paragraph{Visual findings.}  Figure~\ref{fig:qualitative}, described in the Introduction, highlights a visual finding from~\citet{fairman2019marijuana}.  The results are subjectively similar, though a number of relationships may change at the individual level.  Just as authors include visual findings in their papers, we argue that a DP evaluation should include qualitative results as part of an argument for epistemic parity. 

\subsection{Summary of experimental results} 
\label{sec:experiments:summary}

Overall, we were encouraged by the performance of current state-of-the-art synthesizers on our benchmark. DP synthetic data has become more widely used in the social sciences (for Census Data, etc.) and these findings suggest that, in certain contexts, scientists can use DP synthetic data to conduct their scientific inquiry. We caveat this point: \textit{Certain contexts} means relatively low-dimensional tabular data. Our benchmark can be used to assess if those data characteristics hold for a particular dataset, and researchers can proceed with their private analysis with increased confidence.

However, large domain and high-dimensional settings are still a challenge for DP synthesizers: as the domain/number of variables grows, the ease of \textit{fingerprinting} individuals in a dataset increases dramatically. Our findings suggest that existing synthesizers struggle to scale (PrivMRF, MST, AIM, PrivBayes), or are far from achieving reasonable utility (PATECTGAN, GEM). We suggest incorporating more principled methods of data preprocessing, like DP-binning, DP variable pruning, or other domain/variable count reduction techniques into DP synthesizers, so that successful marginals-based methods can be utilized for more complex data.

\section{Conclusions and Future Work}
\label{sec:conc}

\emph{Summary of contributions.} In this paper, we proposed \emph{epistemic parity} as methodology for measuring the utility of DP synthetic data in support of scientific research.  We assembled a benchmark of peer-reviewed papers that analyze one of four studies in the ICPSR social science repository.  We then experimentally evaluated epistemic parity achieved by state-of-the-art DP synthesizers over the papers in our benchmark.  Overall, we found epistemic parity to be a compelling method for evaluating DP synthesizers.  Further, we found that, of the six DP synthesizers we evaluated, no single synthesizer outperformed all others on all papers.  Finally, some findings were never reproduced by any of the synthesizers.

Despite facing well-known reproducibility challenges during benchmark construction, we are confident that our results will lead to generalizable insights. We are continuing to expand the benchmark with additional studies and papers, and invite others to use our open-source extensible framework, \synrd, to contribute. We are also incorporating additional synthesizers, to develop stronger insight into the performance trade-offs between classes of DP synthesizer methods in this setting.

\paragraph{Future work: Characterizing false discoveries.} Replicating published findings using synthetic versions of the original data can reveal some implications of DP for scientific research. However, our epistemic parity methodology does not assess the possibility of findings that \textit{would have occurred} if the original research had been done on synthetic data, or, more generally, the replicability of any hypotheses for which tests were not published. This challenge has been referred to as the file-drawer problem~\cite{rosenthal1979file, iyengar1988selection} or publication bias. It may be a more severe issue when the original dataset contains more variables that could be used to formulate potential hypotheses. In future work, epistemic parity could be extended to quantify the effect of DP noise in producing false discoveries by simulating data with both ``real'' and spurious relationships.

\paragraph{Future work: Rebalancing utility and privacy.} Though DP was developed to provide formal guarantees of privacy with best-effort utility, many practitioners and data providers may want the inverse: strong guarantees of utility with quantifiable, flexible risk of privacy violations that can be managed with policy and accountability rather than mathematical guarantees.  We see this benchmark as promoting a more holistic discussion of socio-technical-legal systems for managing privacy-utility trade-offs. Besides, DP synthesizers, once already trained, can generate arbitrarily large samples at low cost. This makes the power of statistical hypothesis tests, or width of intervals, another concern for scientific research. Epistemic parity could be extended to use for calculating sample sizes, for example by Monte Carlo estimation of the sample size required to achieve a desired power for a particular finding, or for all findings in a given study.

\paragraph{Future work: Improving reproducibility and replicability of scientific discovery.} Replicating published findings using synthetic versions of the original data can reveal some implications of DP for scientific research. However, our methodology does not assess the possibility of findings that \textit{would have occurred} if the original research had been done on synthetic data, or, more generally, the replicability of any hypotheses for which tests were not published. This has been referred to as the file-drawer problem~\cite{rosenthal1979file, iyengar1988selection} or publication bias. It may be a more severe issue when the original dataset contains more variables that could be used to formulate hypotheses. In future work, epistemic parity could be extended to quantify the effect of DP noise in producing findings that would not have been identified from the original data.

\begin{acks}
 This work was supported in part by the National Science Foundation Awards No. 1916505 and 1922658.
\end{acks}
\balance 

\bibliographystyle{ACM-Reference-Format}
\bibliography{sample}


\begin{thebibliography}{67}


\ifx \showCODEN    \undefined \def \showCODEN     #1{\unskip}     \fi
\ifx \showDOI      \undefined \def \showDOI       #1{#1}\fi
\ifx \showISBNx    \undefined \def \showISBNx     #1{\unskip}     \fi
\ifx \showISBNxiii \undefined \def \showISBNxiii  #1{\unskip}     \fi
\ifx \showISSN     \undefined \def \showISSN      #1{\unskip}     \fi
\ifx \showLCCN     \undefined \def \showLCCN      #1{\unskip}     \fi
\ifx \shownote     \undefined \def \shownote      #1{#1}          \fi
\ifx \showarticletitle \undefined \def \showarticletitle #1{#1}   \fi
\ifx \showURL      \undefined \def \showURL       {\relax}        \fi
\providecommand\bibfield[2]{#2}
\providecommand\bibinfo[2]{#2}
\providecommand\natexlab[1]{#1}
\providecommand\showeprint[2][]{arXiv:#2}

\bibitem[\protect\citeauthoryear{Alcobaça, Siqueira, Rivolli, Garcia, Oliva,
  and de~Carvalho}{Alcobaça et~al\mbox{.}}{2020}]%
        {pymfe2020}
\bibfield{author}{\bibinfo{person}{Edesio Alcobaça}, \bibinfo{person}{Felipe
  Siqueira}, \bibinfo{person}{Adriano Rivolli}, \bibinfo{person}{Luís P.~F.
  Garcia}, \bibinfo{person}{Jefferson~T. Oliva}, {and} \bibinfo{person}{André
  C. P. L.~F. de Carvalho}.} \bibinfo{year}{2020}\natexlab{}.
\newblock \showarticletitle{MFE: Towards reproducible meta-feature extraction}.
\newblock \bibinfo{journal}{\emph{Journal of Machine Learning Research}}
  \bibinfo{volume}{21}, \bibinfo{number}{111} (\bibinfo{year}{2020}),
  \bibinfo{pages}{1--5}.
\newblock
\urldef\tempurl%
\url{http://jmlr.org/papers/v21/19-348.html}
\showURL{%
\tempurl}


\bibitem[\protect\citeauthoryear{Assari and Bazargan}{Assari and
  Bazargan}{2019}]%
        {assari2019baseline}
\bibfield{author}{\bibinfo{person}{Shervin Assari} {and}
  \bibinfo{person}{Mohsen Bazargan}.} \bibinfo{year}{2019}\natexlab{}.
\newblock \showarticletitle{Baseline obesity increases 25-year risk of
  mortality due to cerebrovascular disease: role of race}.
\newblock \bibinfo{journal}{\emph{International Journal of Environmental
  Research and Public Health}} \bibinfo{volume}{16}, \bibinfo{number}{19}
  (\bibinfo{year}{2019}), \bibinfo{pages}{3705}.
\newblock


\bibitem[\protect\citeauthoryear{Aydore, Brown, Kearns, Kenthapadi, Melis,
  Roth, and Siva}{Aydore et~al\mbox{.}}{2021}]%
        {aydore2021differentially}
\bibfield{author}{\bibinfo{person}{Sergul Aydore}, \bibinfo{person}{William
  Brown}, \bibinfo{person}{Michael Kearns}, \bibinfo{person}{Krishnaram
  Kenthapadi}, \bibinfo{person}{Luca Melis}, \bibinfo{person}{Aaron Roth},
  {and} \bibinfo{person}{Ankit~A Siva}.} \bibinfo{year}{2021}\natexlab{}.
\newblock \showarticletitle{Differentially private query release through
  adaptive projection}. In \bibinfo{booktitle}{\emph{International Conference
  on Machine Learning}}. PMLR, \bibinfo{pages}{457--467}.
\newblock


\bibitem[\protect\citeauthoryear{Bagdasaryan, Poursaeed, and
  Shmatikov}{Bagdasaryan et~al\mbox{.}}{2019}]%
        {bagdasaryan2019differential}
\bibfield{author}{\bibinfo{person}{Eugene Bagdasaryan}, \bibinfo{person}{Omid
  Poursaeed}, {and} \bibinfo{person}{Vitaly Shmatikov}.}
  \bibinfo{year}{2019}\natexlab{}.
\newblock \showarticletitle{Differential privacy has disparate impact on model
  accuracy}.
\newblock \bibinfo{journal}{\emph{Advances in neural information processing
  systems}}  \bibinfo{volume}{32} (\bibinfo{year}{2019}).
\newblock


\bibitem[\protect\citeauthoryear{Baker}{Baker}{2016}]%
        {baker20161}
\bibfield{author}{\bibinfo{person}{Monya Baker}.}
  \bibinfo{year}{2016}\natexlab{}.
\newblock \showarticletitle{1,500 scientists lift the lid on reproducibility}.
\newblock \bibinfo{journal}{\emph{Nature}} \bibinfo{volume}{533},
  \bibinfo{number}{7604} (\bibinfo{year}{2016}).
\newblock


\bibitem[\protect\citeauthoryear{Barba}{Barba}{2018}]%
        {barba2018terminologies}
\bibfield{author}{\bibinfo{person}{Lorena~A Barba}.}
  \bibinfo{year}{2018}\natexlab{}.
\newblock \showarticletitle{Terminologies for reproducible research}.
\newblock \bibinfo{journal}{\emph{arXiv preprint arXiv:1802.03311}}
  (\bibinfo{year}{2018}).
\newblock


\bibitem[\protect\citeauthoryear{Bowen and Liu}{Bowen and Liu}{2020}]%
        {bowen2020comparative}
\bibfield{author}{\bibinfo{person}{Claire~McKay Bowen} {and}
  \bibinfo{person}{Fang Liu}.} \bibinfo{year}{2020}\natexlab{}.
\newblock \showarticletitle{Comparative study of differentially private data
  synthesis methods}.
\newblock \bibinfo{journal}{\emph{Statist. Sci.}} \bibinfo{volume}{35},
  \bibinfo{number}{2} (\bibinfo{year}{2020}), \bibinfo{pages}{280--307}.
\newblock


\bibitem[\protect\citeauthoryear{Boyd and Sarathy}{Boyd and Sarathy}{2022}]%
        {boyd2022differential}
\bibfield{author}{\bibinfo{person}{Danah Boyd} {and} \bibinfo{person}{Jayshree
  Sarathy}.} \bibinfo{year}{2022}\natexlab{}.
\newblock \showarticletitle{Differential Perspectives: Epistemic Disconnects
  Surrounding the US Census Bureau’s Use of Differential Privacy}.
\newblock \bibinfo{journal}{\emph{Harvard Data Science Review (Forthcoming)}}
  (\bibinfo{year}{2022}).
\newblock


\bibitem[\protect\citeauthoryear{Bun and Steinke}{Bun and Steinke}{2016}]%
        {bun2016concentrated}
\bibfield{author}{\bibinfo{person}{Mark Bun} {and} \bibinfo{person}{Thomas
  Steinke}.} \bibinfo{year}{2016}\natexlab{}.
\newblock \showarticletitle{Concentrated differential privacy: Simplifications,
  extensions, and lower bounds}. In \bibinfo{booktitle}{\emph{Theory of
  Cryptography: 14th International Conference, TCC 2016-B, Beijing, China,
  October 31-November 3, 2016, Proceedings, Part I}}. Springer,
  \bibinfo{pages}{635--658}.
\newblock


\bibitem[\protect\citeauthoryear{Cai, Lei, Wei, and Xiao}{Cai
  et~al\mbox{.}}{2021}]%
        {cai2021data}
\bibfield{author}{\bibinfo{person}{Kuntai Cai}, \bibinfo{person}{Xiaoyu Lei},
  \bibinfo{person}{Jianxin Wei}, {and} \bibinfo{person}{Xiaokui Xiao}.}
  \bibinfo{year}{2021}\natexlab{}.
\newblock \showarticletitle{Data synthesis via differentially private markov
  random fields}.
\newblock \bibinfo{journal}{\emph{Proceedings of the VLDB Endowment}}
  \bibinfo{volume}{14}, \bibinfo{number}{11} (\bibinfo{year}{2021}),
  \bibinfo{pages}{2190--2202}.
\newblock


\bibitem[\protect\citeauthoryear{Camerer, Dreber, Holzmeister, Ho, Huber,
  Johannesson, Kirchler, Nave, Nosek, Pfeiffer, et~al\mbox{.}}{Camerer
  et~al\mbox{.}}{2018}]%
        {camerer2018evaluating}
\bibfield{author}{\bibinfo{person}{Colin~F Camerer}, \bibinfo{person}{Anna
  Dreber}, \bibinfo{person}{Felix Holzmeister}, \bibinfo{person}{Teck-Hua Ho},
  \bibinfo{person}{J{\"u}rgen Huber}, \bibinfo{person}{Magnus Johannesson},
  \bibinfo{person}{Michael Kirchler}, \bibinfo{person}{Gideon Nave},
  \bibinfo{person}{Brian~A Nosek}, \bibinfo{person}{Thomas Pfeiffer},
  {et~al\mbox{.}}} \bibinfo{year}{2018}\natexlab{}.
\newblock \showarticletitle{Evaluating the replicability of social science
  experiments in Nature and Science between 2010 and 2015}.
\newblock \bibinfo{journal}{\emph{Nature Human Behaviour}} \bibinfo{volume}{2},
  \bibinfo{number}{9} (\bibinfo{year}{2018}), \bibinfo{pages}{637--644}.
\newblock


\bibitem[\protect\citeauthoryear{Christ, Radway, and Bellovin}{Christ
  et~al\mbox{.}}{2022}]%
        {christ2022differential}
\bibfield{author}{\bibinfo{person}{Miranda Christ}, \bibinfo{person}{Sarah
  Radway}, {and} \bibinfo{person}{Steven~M Bellovin}.}
  \bibinfo{year}{2022}\natexlab{}.
\newblock \showarticletitle{Differential Privacy and Swapping: Examining
  De-Identification's Impact on Minority Representation and Privacy
  Preservation in the US Census}. In \bibinfo{booktitle}{\emph{2022 IEEE
  Symposium on Security and Privacy (SP)}}. IEEE Computer Society,
  \bibinfo{pages}{1564--1564}.
\newblock


\bibitem[\protect\citeauthoryear{Cohen, Xia, Zweigenbaum, Callahan, Hargraves,
  Goss, Ide, N{\'e}v{\'e}ol, Grouin, and Hunter}{Cohen et~al\mbox{.}}{2018}]%
        {cohen2018three}
\bibfield{author}{\bibinfo{person}{K~Bretonnel Cohen}, \bibinfo{person}{Jingbo
  Xia}, \bibinfo{person}{Pierre Zweigenbaum}, \bibinfo{person}{Tiffany~J
  Callahan}, \bibinfo{person}{Orin Hargraves}, \bibinfo{person}{Foster Goss},
  \bibinfo{person}{Nancy Ide}, \bibinfo{person}{Aur{\'e}lie N{\'e}v{\'e}ol},
  \bibinfo{person}{Cyril Grouin}, {and} \bibinfo{person}{Lawrence~E Hunter}.}
  \bibinfo{year}{2018}\natexlab{}.
\newblock \showarticletitle{Three dimensions of reproducibility in natural
  language processing}. In \bibinfo{booktitle}{\emph{LREC... International
  Conference on Language Resources \& Evaluation:[proceedings]. International
  Conference on Language Resources and Evaluation}},
  Vol.~\bibinfo{volume}{2018}. NIH Public Access, \bibinfo{pages}{156}.
\newblock


\bibitem[\protect\citeauthoryear{Dalton, Ingels, and Fritch}{Dalton
  et~al\mbox{.}}{2016}]%
        {dalton2016high}
\bibfield{author}{\bibinfo{person}{Ben Dalton}, \bibinfo{person}{Steven~J
  Ingels}, {and} \bibinfo{person}{Laura Fritch}.}
  \bibinfo{year}{2016}\natexlab{}.
\newblock \bibinfo{booktitle}{\emph{High School Longitudinal Study of 2009
  (HSLS:09). 2013 Update and High School Transcript Study: A First Look at Fall
  2009 Ninth-Graders in 2013. NCES 2015-037rev.}}
\newblock \bibinfo{type}{{T}echnical {R}eport} ICPSR36423.v1.
  \bibinfo{institution}{Inter-University Consortium for Political and Social
  Research [distributor]}.
\newblock
\newblock
\shownote{https://doi.org/10.3886/ICPSR36423.v1.}


\bibitem[\protect\citeauthoryear{Ding, Hardt, Miller, and Schmidt}{Ding
  et~al\mbox{.}}{2021}]%
        {ding2021retiring}
\bibfield{author}{\bibinfo{person}{Frances Ding}, \bibinfo{person}{Moritz
  Hardt}, \bibinfo{person}{John Miller}, {and} \bibinfo{person}{Ludwig
  Schmidt}.} \bibinfo{year}{2021}\natexlab{}.
\newblock \showarticletitle{Retiring adult: New datasets for fair machine
  learning}.
\newblock \bibinfo{journal}{\emph{NeurIPS}} (\bibinfo{year}{2021}).
\newblock


\bibitem[\protect\citeauthoryear{Ding, Zhang, Li, Wang, Yu, and Pan}{Ding
  et~al\mbox{.}}{2020}]%
        {ding2020differentially}
\bibfield{author}{\bibinfo{person}{Jiahao Ding}, \bibinfo{person}{Xinyue
  Zhang}, \bibinfo{person}{Xiaohuan Li}, \bibinfo{person}{Junyi Wang},
  \bibinfo{person}{Rong Yu}, {and} \bibinfo{person}{Miao Pan}.}
  \bibinfo{year}{2020}\natexlab{}.
\newblock \showarticletitle{Differentially private and fair classification via
  calibrated functional mechanism}. In \bibinfo{booktitle}{\emph{AAAI}},
  Vol.~\bibinfo{volume}{34}. \bibinfo{pages}{622--629}.
\newblock


\bibitem[\protect\citeauthoryear{Dinur and Nissim}{Dinur and Nissim}{2003}]%
        {DBLP:conf/pods/DinurN03}
\bibfield{author}{\bibinfo{person}{Irit Dinur} {and} \bibinfo{person}{Kobbi
  Nissim}.} \bibinfo{year}{2003}\natexlab{}.
\newblock \showarticletitle{Revealing information while preserving privacy}. In
  \bibinfo{booktitle}{\emph{Proceedings of the Twenty-Second {ACM}
  {SIGACT-SIGMOD-SIGART} Symposium on Principles of Database Systems, June
  9-12, 2003, San Diego, CA, {USA}}}, \bibfield{editor}{\bibinfo{person}{Frank
  Neven}, \bibinfo{person}{Catriel Beeri}, {and} \bibinfo{person}{Tova Milo}}
  (Eds.). \bibinfo{publisher}{{ACM}}, \bibinfo{pages}{202--210}.
\newblock
\urldef\tempurl%
\url{https://doi.org/10.1145/773153.773173}
\showDOI{\tempurl}


\bibitem[\protect\citeauthoryear{Dua and Graff}{Dua and Graff}{2017}]%
        {dua2019UCI}
\bibfield{author}{\bibinfo{person}{Dheeru Dua} {and} \bibinfo{person}{Casey
  Graff}.} \bibinfo{year}{2017}\natexlab{}.
\newblock \bibinfo{title}{{UCI} Machine Learning Repository}.
\newblock
\newblock
\urldef\tempurl%
\url{http://archive.ics.uci.edu/ml}
\showURL{%
\tempurl}


\bibitem[\protect\citeauthoryear{Dwork, McSherry, Nissim, and Smith}{Dwork
  et~al\mbox{.}}{2006}]%
        {dwork2006calibrating}
\bibfield{author}{\bibinfo{person}{Cynthia Dwork}, \bibinfo{person}{Frank
  McSherry}, \bibinfo{person}{Kobbi Nissim}, {and} \bibinfo{person}{Adam
  Smith}.} \bibinfo{year}{2006}\natexlab{}.
\newblock \showarticletitle{Calibrating noise to sensitivity in private data
  analysis}. In \bibinfo{booktitle}{\emph{Theory of cryptography conference}}.
  Springer, \bibinfo{pages}{265--284}.
\newblock


\bibitem[\protect\citeauthoryear{Dwork, Naor, Reingold, Rothblum, and
  Vadhan}{Dwork et~al\mbox{.}}{2009}]%
        {dwork2009complexity}
\bibfield{author}{\bibinfo{person}{Cynthia Dwork}, \bibinfo{person}{Moni Naor},
  \bibinfo{person}{Omer Reingold}, \bibinfo{person}{Guy~N Rothblum}, {and}
  \bibinfo{person}{Salil Vadhan}.} \bibinfo{year}{2009}\natexlab{}.
\newblock \showarticletitle{On the complexity of differentially private data
  release: efficient algorithms and hardness results}. In
  \bibinfo{booktitle}{\emph{Proceedings of the forty-first annual ACM symposium
  on Theory of computing}}. \bibinfo{pages}{381--390}.
\newblock


\bibitem[\protect\citeauthoryear{Dwork, Roth, et~al\mbox{.}}{Dwork
  et~al\mbox{.}}{2014}]%
        {dwork2014algorithmic}
\bibfield{author}{\bibinfo{person}{Cynthia Dwork}, \bibinfo{person}{Aaron
  Roth}, {et~al\mbox{.}}} \bibinfo{year}{2014}\natexlab{}.
\newblock \showarticletitle{The algorithmic foundations of differential
  privacy.}
\newblock \bibinfo{journal}{\emph{Foundations and Trends in Theoretical
  Computer Science}} \bibinfo{volume}{9}, \bibinfo{number}{3-4}
  (\bibinfo{year}{2014}), \bibinfo{pages}{211--407}.
\newblock


\bibitem[\protect\citeauthoryear{Errington, Denis, Perfito, Iorns, and
  Nosek}{Errington et~al\mbox{.}}{2021}]%
        {errington2021reproducibility}
\bibfield{author}{\bibinfo{person}{Timothy~M Errington},
  \bibinfo{person}{Alexandria Denis}, \bibinfo{person}{Nicole Perfito},
  \bibinfo{person}{Elizabeth Iorns}, {and} \bibinfo{person}{Brian~A Nosek}.}
  \bibinfo{year}{2021}\natexlab{}.
\newblock \showarticletitle{Reproducibility in cancer biology: challenges for
  assessing replicability in preclinical cancer biology}.
\newblock \bibinfo{journal}{\emph{Elife}}  \bibinfo{volume}{10}
  (\bibinfo{year}{2021}), \bibinfo{pages}{e67995}.
\newblock


\bibitem[\protect\citeauthoryear{Fairman, Furr-Holden, and Johnson}{Fairman
  et~al\mbox{.}}{2019}]%
        {fairman2019marijuana}
\bibfield{author}{\bibinfo{person}{Brian~J Fairman}, \bibinfo{person}{C~Debra
  Furr-Holden}, {and} \bibinfo{person}{Renee~M Johnson}.}
  \bibinfo{year}{2019}\natexlab{}.
\newblock \showarticletitle{When marijuana is used before cigarettes or
  alcohol: Demographic predictors and associations with heavy use, cannabis use
  disorder, and other drug-related outcomes}.
\newblock \bibinfo{journal}{\emph{Prevention Science}} \bibinfo{volume}{20},
  \bibinfo{number}{2} (\bibinfo{year}{2019}), \bibinfo{pages}{225--233}.
\newblock


\bibitem[\protect\citeauthoryear{Fruiht and Chan}{Fruiht and Chan}{2018}]%
        {fruiht2018naturally}
\bibfield{author}{\bibinfo{person}{Veronica Fruiht} {and}
  \bibinfo{person}{Thomas Chan}.} \bibinfo{year}{2018}\natexlab{}.
\newblock \showarticletitle{Naturally {{Occurring Mentorship}} in a {{National
  Sample}} of {{First-Generation College Goers}}: {{A Promising Portal}} for
  {{Academic}} and {{Developmental Success}}}.
\newblock  \bibinfo{volume}{61}, \bibinfo{number}{3-4} (\bibinfo{year}{2018}),
  \bibinfo{pages}{386--397}.
\newblock
\showISSN{00910562}
\urldef\tempurl%
\url{https://doi.org/10.1002/ajcp.12233}
\showDOI{\tempurl}


\bibitem[\protect\citeauthoryear{Ganev, Oprisanu, and De~Cristofaro}{Ganev
  et~al\mbox{.}}{2021}]%
        {ganev2021robin}
\bibfield{author}{\bibinfo{person}{Georgi Ganev}, \bibinfo{person}{Bristena
  Oprisanu}, {and} \bibinfo{person}{Emiliano De~Cristofaro}.}
  \bibinfo{year}{2021}\natexlab{}.
\newblock \showarticletitle{Robin Hood and Matthew Effects--Differential
  Privacy Has Disparate Impact on Synthetic Data}.
\newblock \bibinfo{journal}{\emph{arXiv preprint arXiv:2109.11429}}
  (\bibinfo{year}{2021}).
\newblock


\bibitem[\protect\citeauthoryear{Hardt, Ligett, and McSherry}{Hardt
  et~al\mbox{.}}{2010}]%
        {hardt2010simple}
\bibfield{author}{\bibinfo{person}{Moritz Hardt}, \bibinfo{person}{Katrina
  Ligett}, {and} \bibinfo{person}{Frank McSherry}.}
  \bibinfo{year}{2010}\natexlab{}.
\newblock \showarticletitle{A simple and practical algorithm for differentially
  private data release}.
\newblock \bibinfo{journal}{\emph{arXiv preprint arXiv:1012.4763}}
  (\bibinfo{year}{2010}).
\newblock


\bibitem[\protect\citeauthoryear{{Harris, Kathleen Mullan} and {Udry, J.
  Richard}}{{Harris, Kathleen Mullan} and {Udry, J. Richard}}{2022}]%
        {addhealth}
\bibfield{author}{\bibinfo{person}{{Harris, Kathleen Mullan}} {and}
  \bibinfo{person}{{Udry, J. Richard}}.} \bibinfo{year}{2022}\natexlab{}.
\newblock \bibinfo{booktitle}{\emph{National Longitudinal Study of Adolescent
  to Adult Health (Add Health), 1994-2018 [Public Use]}}.
\newblock \bibinfo{type}{{T}echnical {R}eport} ICPSR21600.v25.
  \bibinfo{institution}{Inter-university Consortium for Political and Social
  Research [distributor], Carolina Population Center, University of North
  Carolina-Chapel Hill [distributor]}.
\newblock
\newblock
\shownote{https://doi.org/10.3886/ICPSR21600.v25.}


\bibitem[\protect\citeauthoryear{Hay, Machanavajjhala, Miklau, Chen, and
  Zhang}{Hay et~al\mbox{.}}{2016}]%
        {hay2016principled}
\bibfield{author}{\bibinfo{person}{Michael Hay}, \bibinfo{person}{Ashwin
  Machanavajjhala}, \bibinfo{person}{Gerome Miklau}, \bibinfo{person}{Yan
  Chen}, {and} \bibinfo{person}{Dan Zhang}.} \bibinfo{year}{2016}\natexlab{}.
\newblock \showarticletitle{Principled evaluation of differentially private
  algorithms using dpbench}. In \bibinfo{booktitle}{\emph{Proceedings of the
  2016 International Conference on Management of Data}}.
  \bibinfo{pages}{139--154}.
\newblock


\bibitem[\protect\citeauthoryear{Hill}{Hill}{2015}]%
        {hill2015evaluating}
\bibfield{author}{\bibinfo{person}{Raquel Hill}.}
  \bibinfo{year}{2015}\natexlab{}.
\newblock \showarticletitle{Evaluating the Utility of Differential Privacy: A
  Use Case Study of a Behavioral Science Dataset}.
\newblock In \bibinfo{booktitle}{\emph{Medical Data Privacy Handbook}}.
  \bibinfo{publisher}{Springer}, \bibinfo{pages}{59--82}.
\newblock


\bibitem[\protect\citeauthoryear{House}{House}{2018}]%
        {acls}
\bibfield{author}{\bibinfo{person}{James~S. House}.}
  \bibinfo{year}{2018}\natexlab{}.
\newblock \bibinfo{booktitle}{\emph{Americans’ Changing Lives: Waves I, II,
  III, IV, and V, 1986, 1989, 1994, 2002, and 2011}}.
\newblock \bibinfo{type}{{T}echnical {R}eport} ICPSR04690.v9.
  \bibinfo{institution}{Inter-university Consortium for Political and Social
  Research [distributor]}.
\newblock
\newblock
\shownote{https://doi.org/10.3886/ICPSR04690.v9.}


\bibitem[\protect\citeauthoryear{Iverson and Terry}{Iverson and Terry}{2021}]%
        {iverson2021high}
\bibfield{author}{\bibinfo{person}{Grant~L Iverson} {and}
  \bibinfo{person}{Douglas~P Terry}.} \bibinfo{year}{2021}\natexlab{}.
\newblock \showarticletitle{High school football and risk for depression and
  suicidality in adulthood: findings from a national longitudinal study}.
\newblock \bibinfo{journal}{\emph{Frontiers in neurology}}
  \bibinfo{volume}{12} (\bibinfo{year}{2021}).
\newblock


\bibitem[\protect\citeauthoryear{Iyengar and Greenhouse}{Iyengar and
  Greenhouse}{1988}]%
        {iyengar1988selection}
\bibfield{author}{\bibinfo{person}{Satish Iyengar} {and}
  \bibinfo{person}{Joel~B Greenhouse}.} \bibinfo{year}{1988}\natexlab{}.
\newblock \showarticletitle{Selection models and the file drawer problem}.
\newblock \bibinfo{journal}{\emph{Statist. Sci.}} (\bibinfo{year}{1988}),
  \bibinfo{pages}{109--117}.
\newblock


\bibitem[\protect\citeauthoryear{Jagielski, Kearns, Mao, Oprea, Roth,
  Sharifi-Malvajerdi, and Ullman}{Jagielski et~al\mbox{.}}{2019}]%
        {jagielski2019differentially}
\bibfield{author}{\bibinfo{person}{Matthew Jagielski}, \bibinfo{person}{Michael
  Kearns}, \bibinfo{person}{Jieming Mao}, \bibinfo{person}{Alina Oprea},
  \bibinfo{person}{Aaron Roth}, \bibinfo{person}{Saeed Sharifi-Malvajerdi},
  {and} \bibinfo{person}{Jonathan Ullman}.} \bibinfo{year}{2019}\natexlab{}.
\newblock \showarticletitle{Differentially private fair learning}. In
  \bibinfo{booktitle}{\emph{ICML}}. \bibinfo{pages}{3000--3008}.
\newblock


\bibitem[\protect\citeauthoryear{Jayaraman and Evans}{Jayaraman and
  Evans}{2019}]%
        {jayaraman2019evaluating}
\bibfield{author}{\bibinfo{person}{Bargav Jayaraman} {and}
  \bibinfo{person}{David Evans}.} \bibinfo{year}{2019}\natexlab{}.
\newblock \showarticletitle{Evaluating differentially private machine learning
  in practice}. In \bibinfo{booktitle}{\emph{28th USENIX Security Symposium
  (USENIX Security 19)}}. \bibinfo{pages}{1895--1912}.
\newblock


\bibitem[\protect\citeauthoryear{Jeong, Wu, Dasgupta, Médard, and
  Calmon}{Jeong et~al\mbox{.}}{2021}]%
        {jeong2021}
\bibfield{author}{\bibinfo{person}{Haewon Jeong}, \bibinfo{person}{Michael~D.
  Wu}, \bibinfo{person}{Nilanjana Dasgupta}, \bibinfo{person}{Muriel Médard},
  {and} \bibinfo{person}{Flavio~P. Calmon}.} \bibinfo{year}{2021}\natexlab{}.
\newblock \showarticletitle{WhoGets the Benefit of the Doubt? Racial Bias in
  Machine Learning Algorithms Applied to Secondary School Math Education}. In
  \bibinfo{booktitle}{\emph{{35th Conference on Neural Information Processing
  Systems (NeurIPS 2021) Workshop on Math AI for Education (MATHAI4ED)}}}.
\newblock


\bibitem[\protect\citeauthoryear{Joanes and Gill}{Joanes and Gill}{1998}]%
        {joanes1998ComparingMO}
\bibfield{author}{\bibinfo{person}{Derrick~N. Joanes} {and}
  \bibinfo{person}{Christine~A. Gill}.} \bibinfo{year}{1998}\natexlab{}.
\newblock \showarticletitle{Comparing measures of sample skewness and
  kurtosis}.
\newblock \bibinfo{journal}{\emph{The Statistician}}  \bibinfo{volume}{47}
  (\bibinfo{year}{1998}), \bibinfo{pages}{183--189}.
\newblock


\bibitem[\protect\citeauthoryear{Kenny, Kuriwaki, McCartan, Rosenman, Simko,
  and Imai}{Kenny et~al\mbox{.}}{2021}]%
        {kenny2021use}
\bibfield{author}{\bibinfo{person}{Christopher~T Kenny}, \bibinfo{person}{Shiro
  Kuriwaki}, \bibinfo{person}{Cory McCartan}, \bibinfo{person}{Evan~TR
  Rosenman}, \bibinfo{person}{Tyler Simko}, {and} \bibinfo{person}{Kosuke
  Imai}.} \bibinfo{year}{2021}\natexlab{}.
\newblock \showarticletitle{The use of differential privacy for census data and
  its impact on redistricting: The case of the 2020 US Census}.
\newblock \bibinfo{journal}{\emph{Science advances}} \bibinfo{volume}{7},
  \bibinfo{number}{41} (\bibinfo{year}{2021}), \bibinfo{pages}{eabk3283}.
\newblock


\bibitem[\protect\citeauthoryear{Kohavi and Becker}{Kohavi and Becker}{1996}]%
        {adult}
\bibfield{author}{\bibinfo{person}{Ronny Kohavi} {and} \bibinfo{person}{Barry
  Becker}.} \bibinfo{year}{1996}\natexlab{}.
\newblock \bibinfo{booktitle}{\emph{{UCI} Mushroom Data Set}}.
\newblock \bibinfo{type}{{T}echnical {R}eport}. \bibinfo{institution}{{UCI}
  Machine Learning Repository}.
\newblock
\newblock
\shownote{\url{https://archive.ics.uci.edu/ml/datasets/adult}.}


\bibitem[\protect\citeauthoryear{Lee and Simpkins}{Lee and Simpkins}{2021}]%
        {lee2021ability}
\bibfield{author}{\bibinfo{person}{Glona Lee} {and} \bibinfo{person}{Sandra~D
  Simpkins}.} \bibinfo{year}{2021}\natexlab{}.
\newblock \showarticletitle{Ability self-concepts and parental support may
  protect adolescents when they experience low support from their math
  teachers}.
\newblock \bibinfo{journal}{\emph{Journal of Adolescence}}
  \bibinfo{volume}{88} (\bibinfo{year}{2021}), \bibinfo{pages}{48--57}.
\newblock


\bibitem[\protect\citeauthoryear{Liu, Vietri, and Wu}{Liu
  et~al\mbox{.}}{2021}]%
        {liu2021iterative}
\bibfield{author}{\bibinfo{person}{Terrance Liu}, \bibinfo{person}{Giuseppe
  Vietri}, {and} \bibinfo{person}{Steven~Z Wu}.}
  \bibinfo{year}{2021}\natexlab{}.
\newblock \showarticletitle{Iterative methods for private synthetic data:
  Unifying framework and new methods}.
\newblock \bibinfo{journal}{\emph{Advances in Neural Information Processing
  Systems}}  \bibinfo{volume}{34} (\bibinfo{year}{2021}),
  \bibinfo{pages}{690--702}.
\newblock


\bibitem[\protect\citeauthoryear{McKenna, Miklau, Hay, and
  Machanavajjhala}{McKenna et~al\mbox{.}}{2018}]%
        {mckenna2018optimizing}
\bibfield{author}{\bibinfo{person}{Ryan McKenna}, \bibinfo{person}{Gerome
  Miklau}, \bibinfo{person}{Michael Hay}, {and} \bibinfo{person}{Ashwin
  Machanavajjhala}.} \bibinfo{year}{2018}\natexlab{}.
\newblock \showarticletitle{Optimizing error of high-dimensional statistical
  queries under differential privacy}.
\newblock \bibinfo{journal}{\emph{arXiv preprint arXiv:1808.03537}}
  (\bibinfo{year}{2018}).
\newblock


\bibitem[\protect\citeauthoryear{McKenna, Miklau, and Sheldon}{McKenna
  et~al\mbox{.}}{2021}]%
        {mckenna2021winning}
\bibfield{author}{\bibinfo{person}{Ryan McKenna}, \bibinfo{person}{Gerome
  Miklau}, {and} \bibinfo{person}{Daniel Sheldon}.}
  \bibinfo{year}{2021}\natexlab{}.
\newblock \showarticletitle{Winning the {NIST} Contest: A scalable and general
  approach to differentially private synthetic data}.
\newblock \bibinfo{journal}{\emph{arXiv preprint arXiv:2108.04978}}
  (\bibinfo{year}{2021}).
\newblock


\bibitem[\protect\citeauthoryear{McKenna, Mullins, Sheldon, and Miklau}{McKenna
  et~al\mbox{.}}{2022}]%
        {mckenna2022aim}
\bibfield{author}{\bibinfo{person}{Ryan McKenna}, \bibinfo{person}{Brett
  Mullins}, \bibinfo{person}{Daniel Sheldon}, {and} \bibinfo{person}{Gerome
  Miklau}.} \bibinfo{year}{2022}\natexlab{}.
\newblock \showarticletitle{AIM: An Adaptive and Iterative Mechanism for
  Differentially Private Synthetic Data}.
\newblock \bibinfo{journal}{\emph{arXiv preprint arXiv:2201.12677}}
  (\bibinfo{year}{2022}).
\newblock


\bibitem[\protect\citeauthoryear{Munaf{\`o}, Nosek, Bishop, Button, Chambers,
  Percie~du Sert, Simonsohn, Wagenmakers, Ware, and Ioannidis}{Munaf{\`o}
  et~al\mbox{.}}{2017}]%
        {munafo2017manifesto}
\bibfield{author}{\bibinfo{person}{Marcus~R Munaf{\`o}},
  \bibinfo{person}{Brian~A Nosek}, \bibinfo{person}{Dorothy~VM Bishop},
  \bibinfo{person}{Katherine~S Button}, \bibinfo{person}{Christopher~D
  Chambers}, \bibinfo{person}{Nathalie Percie~du Sert}, \bibinfo{person}{Uri
  Simonsohn}, \bibinfo{person}{Eric-Jan Wagenmakers},
  \bibinfo{person}{Jennifer~J Ware}, {and} \bibinfo{person}{John Ioannidis}.}
  \bibinfo{year}{2017}\natexlab{}.
\newblock \showarticletitle{A manifesto for reproducible science}.
\newblock \bibinfo{journal}{\emph{Nature human behaviour}} \bibinfo{volume}{1},
  \bibinfo{number}{1} (\bibinfo{year}{2017}), \bibinfo{pages}{1--9}.
\newblock


\bibitem[\protect\citeauthoryear{{National Academies of Sciences, Engineering,
  and Medicine and others}}{{National Academies of Sciences, Engineering, and
  Medicine and others}}{2019}]%
        {national2019reproducibility}
\bibfield{author}{\bibinfo{person}{{National Academies of Sciences,
  Engineering, and Medicine and others}}.} \bibinfo{year}{2019}\natexlab{}.
\newblock \showarticletitle{Reproducibility and replicability in science}.
\newblock  (\bibinfo{year}{2019}).
\newblock


\bibitem[\protect\citeauthoryear{Nosek, Ebersole, DeHaven, and Mellor}{Nosek
  et~al\mbox{.}}{2018}]%
        {nosek2018preregistration}
\bibfield{author}{\bibinfo{person}{Brian~A Nosek}, \bibinfo{person}{Charles~R
  Ebersole}, \bibinfo{person}{Alexander~C DeHaven}, {and}
  \bibinfo{person}{David~T Mellor}.} \bibinfo{year}{2018}\natexlab{}.
\newblock \showarticletitle{The preregistration revolution}.
\newblock \bibinfo{journal}{\emph{Proceedings of the National Academy of
  Sciences}} \bibinfo{volume}{115}, \bibinfo{number}{11}
  (\bibinfo{year}{2018}), \bibinfo{pages}{2600--2606}.
\newblock


\bibitem[\protect\citeauthoryear{Pierce and Quiroz}{Pierce and Quiroz}{2019}]%
        {PierceQuiroz2019}
\bibfield{author}{\bibinfo{person}{Kayla D.~R. Pierce} {and}
  \bibinfo{person}{Christopher~S. Quiroz}.} \bibinfo{year}{2019}\natexlab{}.
\newblock \showarticletitle{Who matters most? Social support, social strain,
  and emotions}.
\newblock \bibinfo{journal}{\emph{Journal of Social and Personal
  Relationships}} \bibinfo{volume}{36}, \bibinfo{number}{10}
  (\bibinfo{year}{2019}), \bibinfo{pages}{3273–3292}.
\newblock


\bibitem[\protect\citeauthoryear{Ping, Stoyanovich, and Howe}{Ping
  et~al\mbox{.}}{2017}]%
        {DBLP:conf/ssdbm/PingSH17}
\bibfield{author}{\bibinfo{person}{Haoyue Ping}, \bibinfo{person}{Julia
  Stoyanovich}, {and} \bibinfo{person}{Bill Howe}.}
  \bibinfo{year}{2017}\natexlab{}.
\newblock \showarticletitle{DataSynthesizer: Privacy-Preserving Synthetic
  Datasets}. In \bibinfo{booktitle}{\emph{Proceedings of the 29th International
  Conference on Scientific and Statistical Database Management, Chicago, IL,
  USA, June 27-29, 2017}}. \bibinfo{publisher}{{ACM}},
  \bibinfo{pages}{42:1--42:5}.
\newblock
\urldef\tempurl%
\url{https://doi.org/10.1145/3085504.3091117}
\showDOI{\tempurl}


\bibitem[\protect\citeauthoryear{Pinto, Soares, and Mendes-Moreira}{Pinto
  et~al\mbox{.}}{2016}]%
        {pinto2016towards}
\bibfield{author}{\bibinfo{person}{F{\'a}bio Pinto}, \bibinfo{person}{Carlos
  Soares}, {and} \bibinfo{person}{Jo{\~a}o Mendes-Moreira}.}
  \bibinfo{year}{2016}\natexlab{}.
\newblock \showarticletitle{Towards Automatic Generation of Metafeatures}. In
  \bibinfo{booktitle}{\emph{Advances in Knowledge Discovery and Data Mining}},
  \bibfield{editor}{\bibinfo{person}{James Bailey}, \bibinfo{person}{Latifur
  Khan}, \bibinfo{person}{Takashi Washio}, \bibinfo{person}{Gill Dobbie},
  \bibinfo{person}{Joshua~Zhexue Huang}, {and} \bibinfo{person}{Ruili Wang}}
  (Eds.). \bibinfo{publisher}{Springer International Publishing},
  \bibinfo{address}{Cham}, \bibinfo{pages}{215--226}.
\newblock


\bibitem[\protect\citeauthoryear{Preacher and Hayes}{Preacher and
  Hayes}{2008}]%
        {preacher2008asymptotic}
\bibfield{author}{\bibinfo{person}{Kristopher~J. Preacher} {and}
  \bibinfo{person}{Andrew~F. Hayes}.} \bibinfo{year}{2008}\natexlab{}.
\newblock \showarticletitle{Asymptotic and Resampling Strategies for Assessing
  and Comparing Indirect Effects in Multiple Mediator Models}.
\newblock  \bibinfo{volume}{40}, \bibinfo{number}{3} (\bibinfo{year}{2008}),
  \bibinfo{pages}{879--891}.
\newblock
\showISSN{1554-3528}
\urldef\tempurl%
\url{https://doi.org/10.3758/BRM.40.3.879}
\showDOI{\tempurl}


\bibitem[\protect\citeauthoryear{Raghunathan, Reiter, and Rubin}{Raghunathan
  et~al\mbox{.}}{2003}]%
        {raghunathan2003multiple}
\bibfield{author}{\bibinfo{person}{Trivellore~E Raghunathan},
  \bibinfo{person}{Jerome~P Reiter}, {and} \bibinfo{person}{Donald~B Rubin}.}
  \bibinfo{year}{2003}\natexlab{}.
\newblock \showarticletitle{Multiple imputation for statistical disclosure
  limitation}.
\newblock \bibinfo{journal}{\emph{Journal of official statistics}}
  \bibinfo{volume}{19}, \bibinfo{number}{1} (\bibinfo{year}{2003}),
  \bibinfo{pages}{1}.
\newblock


\bibitem[\protect\citeauthoryear{R{\"a}is{\"a}, J{\"a}lk{\"o}, Kaski, and
  Honkela}{R{\"a}is{\"a} et~al\mbox{.}}{2022}]%
        {raisa2022noise}
\bibfield{author}{\bibinfo{person}{Ossi R{\"a}is{\"a}}, \bibinfo{person}{Joonas
  J{\"a}lk{\"o}}, \bibinfo{person}{Samuel Kaski}, {and} \bibinfo{person}{Antti
  Honkela}.} \bibinfo{year}{2022}\natexlab{}.
\newblock \showarticletitle{Noise-Aware Statistical Inference with
  Differentially Private Synthetic Data}.
\newblock \bibinfo{journal}{\emph{arXiv preprint arXiv:2205.14485}}
  (\bibinfo{year}{2022}).
\newblock


\bibitem[\protect\citeauthoryear{Rivolli, Garcia, Soares, Vanschoren, and
  de~Leon Ferreira~de Carvalho}{Rivolli et~al\mbox{.}}{2018}]%
        {Rivolli2018CharacterizingCD}
\bibfield{author}{\bibinfo{person}{Adriano Rivolli}, \bibinfo{person}{Lu{\'i}s
  Paulo~F. Garcia}, \bibinfo{person}{Carlos Soares}, \bibinfo{person}{Joaquin
  Vanschoren}, {and} \bibinfo{person}{Andr{\'e} Carlos~Ponce de~Leon
  Ferreira~de Carvalho}.} \bibinfo{year}{2018}\natexlab{}.
\newblock \showarticletitle{Characterizing classification datasets: a study of
  meta-features for meta-learning.}
\newblock \bibinfo{journal}{\emph{arXiv: Learning}} (\bibinfo{year}{2018}).
\newblock


\bibitem[\protect\citeauthoryear{Rosenblatt, Allen, and Stoyanovich}{Rosenblatt
  et~al\mbox{.}}{2022}]%
        {DBLP:journals/corr/abs-2204-12903}
\bibfield{author}{\bibinfo{person}{Lucas Rosenblatt}, \bibinfo{person}{Joshua
  Allen}, {and} \bibinfo{person}{Julia Stoyanovich}.}
  \bibinfo{year}{2022}\natexlab{}.
\newblock \showarticletitle{Spending Privacy Budget Fairly and Wisely}.
\newblock \bibinfo{journal}{\emph{CoRR}}  \bibinfo{volume}{abs/2204.12903}
  (\bibinfo{year}{2022}).
\newblock
\urldef\tempurl%
\url{https://doi.org/10.48550/arXiv.2204.12903}
\showDOI{\tempurl}
\showeprint[arXiv]{2204.12903}


\bibitem[\protect\citeauthoryear{Rosenblatt, Liu, Pouyanfar, de~Leon, Desai,
  and Allen}{Rosenblatt et~al\mbox{.}}{2020}]%
        {rosenblatt2020differentially}
\bibfield{author}{\bibinfo{person}{Lucas Rosenblatt}, \bibinfo{person}{Xiaoyan
  Liu}, \bibinfo{person}{Samira Pouyanfar}, \bibinfo{person}{Eduardo de Leon},
  \bibinfo{person}{Anuj Desai}, {and} \bibinfo{person}{Joshua Allen}.}
  \bibinfo{year}{2020}\natexlab{}.
\newblock \showarticletitle{Differentially private synthetic data: Applied
  evaluations and enhancements}.
\newblock \bibinfo{journal}{\emph{arXiv preprint arXiv:2011.05537}}
  (\bibinfo{year}{2020}).
\newblock


\bibitem[\protect\citeauthoryear{Rosenthal}{Rosenthal}{1979}]%
        {rosenthal1979file}
\bibfield{author}{\bibinfo{person}{Robert Rosenthal}.}
  \bibinfo{year}{1979}\natexlab{}.
\newblock \showarticletitle{The file drawer problem and tolerance for null
  results.}
\newblock \bibinfo{journal}{\emph{Psychological bulletin}}
  \bibinfo{volume}{86}, \bibinfo{number}{3} (\bibinfo{year}{1979}),
  \bibinfo{pages}{638}.
\newblock


\bibitem[\protect\citeauthoryear{Rubin}{Rubin}{1981}]%
        {rubin1981bayesian}
\bibfield{author}{\bibinfo{person}{Donald~B Rubin}.}
  \bibinfo{year}{1981}\natexlab{}.
\newblock \showarticletitle{The bayesian bootstrap}.
\newblock \bibinfo{journal}{\emph{The annals of statistics}}
  (\bibinfo{year}{1981}), \bibinfo{pages}{130--134}.
\newblock


\bibitem[\protect\citeauthoryear{Ruggles, Fitch, Magnuson, and
  Schroeder}{Ruggles et~al\mbox{.}}{2019}]%
        {ruggles2019differential}
\bibfield{author}{\bibinfo{person}{Steven Ruggles}, \bibinfo{person}{Catherine
  Fitch}, \bibinfo{person}{Diana Magnuson}, {and} \bibinfo{person}{Jonathan
  Schroeder}.} \bibinfo{year}{2019}\natexlab{}.
\newblock \showarticletitle{Differential privacy and census data: Implications
  for social and economic research}. In \bibinfo{booktitle}{\emph{AEA papers
  and proceedings}}, Vol.~\bibinfo{volume}{109}. \bibinfo{pages}{403--08}.
\newblock


\bibitem[\protect\citeauthoryear{Saw, Chang, and Chan}{Saw
  et~al\mbox{.}}{2018}]%
        {saw2018cross}
\bibfield{author}{\bibinfo{person}{Guan Saw}, \bibinfo{person}{Chi-Ning Chang},
  {and} \bibinfo{person}{Hsun-Yu Chan}.} \bibinfo{year}{2018}\natexlab{}.
\newblock \showarticletitle{Cross-sectional and longitudinal disparities in
  STEM career aspirations at the intersection of gender, race/ethnicity, and
  socioeconomic status}.
\newblock \bibinfo{journal}{\emph{Educational Researcher}}
  \bibinfo{volume}{47}, \bibinfo{number}{8} (\bibinfo{year}{2018}),
  \bibinfo{pages}{525--531}.
\newblock


\bibitem[\protect\citeauthoryear{Schlimmer}{Schlimmer}{1987}]%
        {mushroom}
\bibfield{author}{\bibinfo{person}{Jeff Schlimmer}.}
  \bibinfo{year}{1987}\natexlab{}.
\newblock \bibinfo{booktitle}{\emph{{UCI} Adult Data Set}}.
\newblock \bibinfo{type}{{T}echnical {R}eport}. \bibinfo{institution}{{UCI}
  Machine Learning Repository}.
\newblock
\newblock
\shownote{\url{https://archive.ics.uci.edu/ml/datasets/mushroom}.}


\bibitem[\protect\citeauthoryear{Takagi, Takahashi, Cao, and Yoshikawa}{Takagi
  et~al\mbox{.}}{2021}]%
        {takagi2021p3gm}
\bibfield{author}{\bibinfo{person}{Shun Takagi}, \bibinfo{person}{Tsubasa
  Takahashi}, \bibinfo{person}{Yang Cao}, {and} \bibinfo{person}{Masatoshi
  Yoshikawa}.} \bibinfo{year}{2021}\natexlab{}.
\newblock \showarticletitle{P3gm: Private high-dimensional data release via
  privacy preserving phased generative model}. In
  \bibinfo{booktitle}{\emph{2021 IEEE 37th International Conference on Data
  Engineering (ICDE)}}. IEEE, \bibinfo{pages}{169--180}.
\newblock


\bibitem[\protect\citeauthoryear{Tao, McKenna, Hay, Machanavajjhala, and
  Miklau}{Tao et~al\mbox{.}}{2021}]%
        {tao2021benchmarking}
\bibfield{author}{\bibinfo{person}{Yuchao Tao}, \bibinfo{person}{Ryan McKenna},
  \bibinfo{person}{Michael Hay}, \bibinfo{person}{Ashwin Machanavajjhala},
  {and} \bibinfo{person}{Gerome Miklau}.} \bibinfo{year}{2021}\natexlab{}.
\newblock \showarticletitle{Benchmarking Differentially Private Synthetic Data
  Generation Algorithms}.
\newblock \bibinfo{journal}{\emph{arXiv preprint arXiv:2112.09238}}
  (\bibinfo{year}{2021}).
\newblock


\bibitem[\protect\citeauthoryear{Torkzadehmahani, Kairouz, and
  Paten}{Torkzadehmahani et~al\mbox{.}}{2020}]%
        {DBLP:journals/corr/abs-2001-09700}
\bibfield{author}{\bibinfo{person}{Reihaneh Torkzadehmahani},
  \bibinfo{person}{Peter Kairouz}, {and} \bibinfo{person}{Benedict Paten}.}
  \bibinfo{year}{2020}\natexlab{}.
\newblock \showarticletitle{{DP-CGAN:} Differentially Private Synthetic Data
  and Label Generation}.
\newblock \bibinfo{journal}{\emph{CoRR}}  \bibinfo{volume}{abs/2001.09700}
  (\bibinfo{year}{2020}).
\newblock
\showeprint[arXiv]{2001.09700}
\urldef\tempurl%
\url{https://arxiv.org/abs/2001.09700}
\showURL{%
\tempurl}


\bibitem[\protect\citeauthoryear{{United States Department of Health and Human
  Services}}{{United States Department of Health and Human Services}}{2016}]%
        {nsfduh}
\bibfield{author}{\bibinfo{person}{{United States Department of Health and
  Human Services}}.} \bibinfo{year}{2016}\natexlab{}.
\newblock \bibinfo{booktitle}{\emph{National Survey on Drug Use and Health
  (NSDUH), 2014}}.
\newblock \bibinfo{type}{{T}echnical {R}eport} ICPSR36361.v1.
  \bibinfo{institution}{Inter-university Consortium for Political and Social
  Research [distributor]}.
\newblock
\newblock
\shownote{https://doi.org/10.3886/ICPSR36361.v1.}


\bibitem[\protect\citeauthoryear{Vietri, Tian, Bun, Steinke, and Wu}{Vietri
  et~al\mbox{.}}{2020}]%
        {vietri2020new}
\bibfield{author}{\bibinfo{person}{Giuseppe Vietri}, \bibinfo{person}{Grace
  Tian}, \bibinfo{person}{Mark Bun}, \bibinfo{person}{Thomas Steinke}, {and}
  \bibinfo{person}{Steven Wu}.} \bibinfo{year}{2020}\natexlab{}.
\newblock \showarticletitle{New oracle-efficient algorithms for private
  synthetic data release}. In \bibinfo{booktitle}{\emph{ICML}}.
  \bibinfo{pages}{9765--9774}.
\newblock


\bibitem[\protect\citeauthoryear{Xu, Skoularidou, Cuesta-Infante, and
  Veeramachaneni}{Xu et~al\mbox{.}}{2019}]%
        {xu2019modeling}
\bibfield{author}{\bibinfo{person}{Lei Xu}, \bibinfo{person}{Maria
  Skoularidou}, \bibinfo{person}{Alfredo Cuesta-Infante}, {and}
  \bibinfo{person}{Kalyan Veeramachaneni}.} \bibinfo{year}{2019}\natexlab{}.
\newblock \showarticletitle{Modeling tabular data using conditional gan}.
\newblock \bibinfo{journal}{\emph{Advances in Neural Information Processing
  Systems}}  \bibinfo{volume}{32} (\bibinfo{year}{2019}).
\newblock


\bibitem[\protect\citeauthoryear{Zhang, Cormode, Procopiuc, Srivastava, and
  Xiao}{Zhang et~al\mbox{.}}{2014}]%
        {zhang2017privbayes}
\bibfield{author}{\bibinfo{person}{Jun Zhang}, \bibinfo{person}{Graham
  Cormode}, \bibinfo{person}{Cecilia~M Procopiuc}, \bibinfo{person}{Divesh
  Srivastava}, {and} \bibinfo{person}{Xiaokui Xiao}.} \bibinfo{year}{SIGMOD
  2014}\natexlab{}.
\newblock \showarticletitle{PrivBayes: Private Data Release via Bayesian
  Networks}.
\newblock  (\bibinfo{year}{SIGMOD 2014}).
\newblock


\end{thebibliography}

\end{document}